\documentclass[prd,aps,10pt,longbibliography,reprint]{revtex4-1}
\usepackage{amsmath,amssymb,bm,bbm}

\newcommand{\eq}{\begin{equation}}
\newcommand{\qe}{\end{equation}}
\newcommand{\eqa}{\begin{eqnarray}}
\newcommand{\qea}{\end{eqnarray}}
\newcommand{\non}{\nonumber}

\newcommand{\al}{\alpha}
\newcommand{\bt}{\beta}

\newcommand{\dd}[2]{\frac{\partial #1}{\partial #2}}
\renewcommand{\d}{\mathrm{d}} 

\newcommand{\ad}{a^\dagger}

\newcommand{\ket}[1]{\left| \! \right. #1\left. \! \right\rangle}
\newcommand{\bra}[1]{\left\langle \! \right. #1\left.\! \right| }
\newcommand{\kket}[1]{\!\parallel\! #1\left. \! \right\rangle}
\newcommand{\bbra}[1]{\left\langle \! \right. #1\! \parallel\! }
\newcommand{\ketbra}[1]{\ket{#1}\bra{#1}}
\newcommand{\braket}[2]{\left\langle \! \right. #1\left.\! \right|\! #2\left. \! \right\rangle}

\newcommand{\comment}[1]{}

\begin{document}
\title{
Noncommutative Spherically Symmetric Spaces}

\author{Se\'an Murray}
\email{sean.murray@uclouvain.be}
\affiliation{Centre for Cosmology, Particle Physics and Phenomenology, Universit\'e catholique de Louvain, Chemin du Cyclotron 2, B-1348 Louvain-la-Neuve, Belgium}

\author{Jan Govaerts}
\email{jan.govaerts@uclouvain.be}
\thanks{Fellow of the Stellenbosch Institute for Advanced Study (STIAS), 7600 Stellenbosch, South Africa.}

\affiliation{Centre for Cosmology, Particle Physics and Phenomenology, Universit\'e catholique de Louvain, Chemin du Cyclotron 2, B-1348 Louvain-la-Neuve, Belgium}
\affiliation{International Chair in Mathematical Physics and Applications, University of Abomey-Calavi, 072 B. P. 50, Cotonou, Republic of Benin}

\begin{abstract}We examine some noncommutative spherically symmetric spaces in three space dimensions. A generalization of Snyder's noncommutative (Euclidean) space allows the inclusion of the generator of dilations into the defining algebra of the coordinate and rotation operators. We then construct a spherically symmetric noncommutative Laplacian on this space having the correct limiting spectrum. This is presented via a creation and annihilation operator realization of the algebra, which may lend itself to a truncation of the Hilbert space.
\end{abstract}



\maketitle


\section{Introduction}
Over the last two decades or so noncommutativity of space coordinates has become a much persued avenue of research \cite{Connes:1994yd} and is widely considered a promising candidate for physics beyond the Planck scale \cite{Doplicher:1994zv,Doplicher:1994tu,Douglas:2001ba}. As well as its use in regularization of quantum field theories, noncommuting coordinates have also appeared naturally within string theory \cite{Seiberg:1999vs,Myers:1999ps} and in physical systems with strong magnetic fields \cite{Dunne:1992ew,Bander:2004nj,Frenkel:2004ff,Govaerts:2009ri}.

A closely related development is that of fuzzy physics (see \cite{Balachandran:2005ew} for a review of certain aspects) where finite matrix algebras are used to approximate the algebra of functions on a manifold. Action functionals built from these matrix algebras provide an alternative to lattice actions in the regularization of field theories and are especially natural for field theories on noncommutative spaces. The archetypical fuzzy space is the fuzzy sphere \cite{Berezin:1974du,Hoppe:1982,Madore:1991bw,Grosse:1995jt} but many other spaces have been studied such as the fuzzy complex quadrics \cite{Dolan:2003th}, fuzzy complex projective spaces \cite{Balachandran:2001dd,Dolan:2006tx,Dolan:2007} and the fuzzy Grassmannians \cite{Dolan:2001mi,Murray:2006pi}. Any manifold which can be generated as the coadjoint orbit of a compact Lie group should have a fuzzy description but it is not a necessary condition \cite{Arnlind:2006ux}. For example, Lizzi et al{.} \cite{Lizzi:2003ru,Lizzi:2005zx} have described the fuzzy disc and its spectrum in some detail (see also \cite{Scholtz:2007ig}). The construction relies upon a projection operator, which is applied to the algebra of functions of the noncommutative plane and for which, the creation and annihilation operator description of the NC plane allows a particularly simple form.

In this note, we will present the first step in generalising this to three dimensional space by using a creation and annihilation operators description of a spherically symmetric noncommutative space and an appropriate Laplacian. Hammou et al.~\cite{Hammou:2001cc} have also used an oscillator and projection operator construction to describe a foliation of fuzzy spheres and a discrete radial derivative. Moreno \cite{Moreno:2005cn} took a similar approach in the context of spherically symmetric monopoles. Here, we will maintain noncommutativity in all directions. See \cite{Buric:2008th} for a discussion of the 4d case in the context of black holes.

After a general discussion of matrix elements in section \ref{tensorop}, a realisation using creation and annihilation operators is presented and in section \ref{snyder}, applied to Snyder's noncommutative space \cite{Snyder:1946qz}. In fact, the realisation describes all spherical symmetric noncommutative spaces defined by Snyder-like commutation relations i.e.~by an algebra containing only the coordinates and the generators of rotations with only the coordinate commutator modified (see equations (\ref{gensnyderalg})).

In section \ref{gensny}, we generalise the algebra and postulate additional commutators with the generator of dilations. The Jacobi identity then requires a particular choice to made, which means that the algebra is isomorphic to the Euclidean algebra $e(3)$. A candidate Laplacian is then presented with the correct continuum spectrum and analogous zero modes. Some of its properties are given.

\section{Spherical tensor operators}
\label{tensorop}
We take for noncommutative coordinates the Hermitian operators, $\hat{X}_i~, i=1,2,3$. Spherical symmetry of the noncommutative space they describe imposes\footnote{Repeated indices are summed over.}
\begin{align} [\hat{J}_i,\,\hat{J}_j]&=i\epsilon_{ijk}\hat{J}_k\label{JJcommutator}\\  [\hat{J}_i,\,\hat{X}_j]&=i\epsilon_{ijk}\hat{X}_k ~,\label{JXcommutator}\end{align}
where $\hat{J}_i~, i=1,2,3$ are the generators of rotations.
This is precisely the definition of a spherical vector operator. Writing $\hat{T}^1_0=\hat{X}_3$ and $\hat{T}^1_{\pm1}=\mp\frac{1}{\sqrt{2}}\hat{X}_\pm=\mp\frac{1}{\sqrt{2}}(\hat{X}_1\pm i\hat{X}_2)$ , the matrix elements of such an operator are given by the Wigner-Eckart theorem \cite{Wybourne} as
\begin{align} \bra{n,j,m}\hat{T}^1_\epsilon \ket{n',j',m'}
=C^{j' 1 j}_{m' \epsilon m}\frac{\bbra{n,j}\hat{T}^1\kket{n',j'}}{\sqrt{2j+1}}~,\label{matrixelement}
\end{align}
up to the unknown functions, the reduced matrix elements, $\bbra{n,j-1}\hat{T}^1\kket{n',j}$ and $\bbra{n,j}\hat{T}^1\kket{n',j}$, which do not depend on the label $\epsilon$, and where 
\eq C^{l'l''l}_{m'm''m}=\braket{l',m';l'',m''}{l,m}
\qe are the Clebsch-Gordan coefficients. The indices $n$ and $n'$ represent any other quantum numbers which may appear on specificing the commutators amongst the coordinates $\hat{X}_i$. However, if we suppose that the algebra closes without any additional generators i.e.~if the remaining commutator takes the form
\eq [\hat{X}_i,\,\hat{X}_j]=i\epsilon_{ijk}(\alpha \hat{J}_k+\beta \hat{X}_k) \qquad \alpha,\,\beta \in \mathbb{R}~, 	\label{genXXcommutator}\qe
then we will see that the representations of the algebra and hence the Hilbert space are labelled by the eigenvalues of two Casimir operators, which we suppress here, and by $j$ and $m$.\footnote{In fact, the algebra relations (\ref{JJcommutator}) and (\ref{JXcommutator}) alone imply that the operators $\hat{J}^2$, $\hat{J}_3$ and $\hat{J}\cdot\hat{X}=\hat{X}\cdot\hat{J}$ mutually commute and so their eigenvalues can be used to label the Hilbert space. Using the additional commutator (\ref{genXXcommutator}), we find that $\hat{X}^2$ also mutually commutes and its eigenvalue is also a label. The two Casimirs are formed from $\hat{J}^2$, $\hat{J}\cdot\hat{X}$ and $\hat{X}^2$.}
\comment{If there are no such other quantum numbersthen, for a general rank $1$ tensor operator $\hat{T}^1_\varepsilon$, we have
\begin{align} \hat{T}^1_\varepsilon\ket{j,m}&=(-1)^{j-m'-1}\left(\begin{array}{ccc}
j-1 & 1 & j \\
-m' & \varepsilon & m
\end{array} \right)\bbra{j-1}\hat{T}^1\kket{j}\ket{j-1,m'}\non\\&+(-1)^{j-m'}\left(\begin{array}{ccc}
j & 1 & j \\
-m' & \varepsilon & m
\end{array} \right)\bbra{j}\hat{T}^1\kket{j}\ket{j,m'}\non\\&+(-1)^{j-m'+1}\left(\begin{array}{ccc}
j+1 & 1 & j \\
-m' & \varepsilon & m
\end{array} \right)\bbra{j+1}\hat{T}^1\kket{j}\ket{j+1,m'}
\end{align}
where $m'=m+\varepsilon$.}
Therefore we have defined the Hilbert space in terms of the orthonormal basis\footnote{For generality, we allow spinor representations.}
\eq \mathcal{F}=\{\ket{j,m}:2j\in\mathbb{N},m=-j\ldots j\}.\qe
The Clebsch-Gordan coefficients tell us that the coordinate operators only raise or lower the eigenvalue $j$ by $1$. Therefore, we can restrict $\mathcal{F}$ to contain only either integer or half-odd-integer values of $j$.

For the fuzzy sphere (see \cite{Balachandran:2005ew} and references therein), the coordinates are taken to be proportional to the generators of $\mathop{\rm SU}(2)$, which is the universal cover of the rotation group and we have $\bbra{j-1}\hat{X}_i\kket{j}\equiv 0$. There is then a finite matrix algebra associated to each irreducible representation of $su(2)$. Without such a simple identification, it is more difficult to arrive at a fuzzy (i.e.~finite matrix) description. To obtain operators that can be represented by finite matrices, we need to restrict ourselves to a subspace of the Hilbert space $\mathcal{F}$:
\eq
\mathcal{F}_{j_{0},j_{1}}=\{\ket{j,m}:\,j=j_0,j_0+1\ldots j_{1},\,m=-j\ldots j\}
\qe
with $2j_0,2j_{1}, j_1-j_0\in \mathbb{N}$.
For generality, we have also included a lower bound $j_{0}$.
For the operators themselves, this either means using a projection
\eq \mathbb{P}=\sum_{j=j_0}^{j_1}\sum_{m=-j}^j\ketbra{j,m} \qe
(and operators $\mathbb{P}\hat{X}\mathbb{P}$) or, equivalently, choosing $\bbra{j-1}\hat{T}^1\kket{j}=0$ for $j=j_0$ and $j_1+1$, the matrix elements of the coordinates and their products being the same in either case. Lizzi et al. used such a procedure to describe the fuzzy disc \cite{Lizzi:2003ru,Lizzi:2005zx}. Using the coherent state picture of the noncommutative plane, they introduced a sequence of projection operators, $\hat{P}^{(N)}$, whose Berezin symbols converge to a step function in the radial coordinate in a certain limit. A sequence of finite algebras can be defined by applying the projectors to the full infinite dimensional algebra, $\hat{\mathcal{A}}^{(N)}=\hat{P}^{(N)}\hat{\mathcal{A}}\hat{P}^{(N)}$. Our goal here is to begin the extension of this procedure by describing a noncommutative algebra built from the three rotationally covariant coordinate operators and the Hilbert space it acts on, in such a way that may allow a similar procedure.

\comment{Alternatively, as we will mention later, the space $\mathcal{F}_{j_{0},j_{1}}$ can also be described as the tensor product space
\begin{align} \mathcal{F}_{j_{0},j_{1}}&=\mathrm{span}\{ \ket{k_1,m_1}\otimes\ket{k_2,m_2}:\,m_1=-k_1\ldots k_1,\,m_2=-k_2\ldots k_2\},
\end{align}
where $k_1=\frac{1}{2}(j_1+j_0)$ and $k_2=\frac{1}{2}(j_1-j_0)$.
In this picture, a projection is unnecessary as any rank $1$ tensor operator can be described in terms of the two sets of angular momentum operators, each acting on one of the kets in the tensor product.
}

Before proceeding, let us first note the following facts. Using the matrix elements (\ref{matrixelement}), we can calculate the `radius' $\hat{R}^2=\hat{X}_1^2+\hat{X}_2^2+\hat{X}_3^2$ to find
\eq \hat{R}^2\ket{j,m}=\frac{1}{2j+1}\left(d^2_j+b^2_j+d^2_{j+1}\right)\ket{j,m}\qe
where $d_j=\bbra{j-1}\hat{T}^1\kket{j}$ and $b_j=\bbra{j}\hat{T}^1\kket{j}$. For the projected case, we would have $d_{j_0}=0$, $d_{j_1+1}=0$.

The angular Laplacian is given by
\eq \hat{\Delta}=\sum_{i=1}^3\big[\hat{J}_i,[\hat{J}_i,\,\cdot\,]\big]=\hat{\mathcal{J}}^2~,\qe where $\hat{\mathcal{J}}_i\hat{X}_j=\hat{J}_i\hat{X}_j-\hat{X}_j\hat{J}_i$ and it is easy to see that $\hat{\Delta}\hat{X}_i=2\hat{X}_i$. Observe that the $\hat{J}_i$ and hence $\hat{\Delta}$ commute with any rank $0$ operator whose value on $\ket{j,m}$ does not depend on $m$,
\eq [\hat{J}_i,\,\hat{U}]=0 \quad\mathrm{if}\quad \hat{U}\ket{j,m}=u(j)\ket{j,m}~.\qe
The operator $\hat{R}^2$ is of this form. Such operators provide a map under left and right multiplication amongst rank 1 tensor operators, such as the coordinates, so that for two rank 0 operators $\hat{U}$ and $\hat{U}'$, $\hat{U}\hat{X}_i \hat{U}'$ is another rank 1 tensor operator. It is then clear that the algebra $\hat{\mathcal{A}}$, generated by the coordinate operators $\hat{X}_i$ can be decomposed into eigenspaces of the angular Laplacian. A general operator $\hat{M}\in\hat{\mathcal{A}}$ can be described by the expansion
\eq \hat{M}
=\sum_{l=0}^\infty \sum_{m=-l}^l\sum_d  c_{d,l,m}\hat{\Psi}^{d}_{lm}~,\label{expansion}\qe where $\hat{\Psi}^{d}_{lm}$ are suitably normalised rank $l$ spherical tensor operators, often called polarization tensors in this context, with $d$ representing any degeneracies and the $c_{d,l,m}$ are coefficients in the expansion. We will describe the operators $\hat{\Psi}^{d}_{lm}$ in a little more detail later.

\section{Oscillator description}
We introduce creation and annihilation operators, $a^\al$ and $\ad_\al$, $\al=1,2$, satisfying $[a^\al,\,\ad_\bt]=\delta^\al_\bt$ and $a^\al\ket{0}=0$~.
The Hilbert space $\mathcal{F}$ may then be described by the Fock space spanned by the normalised states
\eq \ket{n_1,n_2}=\frac{1}{\sqrt{n_1!n_2!}} (\ad_1)^{n_1}(\ad_2)^{n_2}\ket{0}, \quad n_1,\,n_2\in\mathbb{N}~.\qe

The angular momentum operators are given by the Schwinger construction as $\hat{J}_i=\ad_\alpha (\frac{\sigma_i}{2})^\alpha{}_\beta a^\beta$ and a basis better suited to them is given by
\eq \ket{j,m}=\frac{N_{j,m}}{\sqrt{(2j)!}}(\hat{J}_-)^{j-m} (\ad_1)^{2j}\ket{0}\qe
with $j=0,\frac{1}{2}\ldots,~m=-j,\ldots j$ and
where $N_{j,m}=\sqrt{\frac{(j+m)!}{(2j)!(j-m)!}}$ and $\hat{J}_\pm=\hat{J}_1\pm i\hat{J}_2$~. The labels $j$ and $m$ are given by the eigenvalues of $\hat{J}^2$ and $\hat{J}_3$ respectively:
\begin{align} \hat{J}^2\ket{j,m}&= \frac{1}{4}\hat{N}(\hat{N}+2)\ket{j,m}= j(j+1)\ket{j,m}~,\\ \hat{J}_3\ket{j,m}& = m\ket{j,m} ~.\end{align}
We also consider terms quadratic in both the creation and annihilation oscillators and present the following formulae
\begin{align}
\hat{J}_\pm\ket{j,m}&=\sqrt{(j\mp m)(j\pm m+1)}\ket{j,m\pm 1}\label{osc1}\\
\ad_1 \ad_2 \ket{j,m}&=\sqrt{(j+m+1)(j-m+1)} \ket{j+1,m}\\
a^1 a^2 \ket{j,m}&=\sqrt{(j+m)(j-m)} \ket{j-1,m}\\
\ad_1 \ad_1 \ket{j,m}&=\sqrt{(j+m+1)(j+m+2)} \ket{j+1,m+1}\\
\ad_2 \ad_2 \ket{j,m}&=\sqrt{(j-m+1)(j-m+2)} \ket{j+1,m-1}\\
a^1 a^1 \ket{j,m}&=\sqrt{(j+m)(j+m-1)} \ket{j-1,m-1}\\
a^2 a^2 \ket{j,m}&=\sqrt{(j-m)(j-m+1)} \ket{j-1,m+1}\label{osc7}~.
\end{align}
\comment{We need to include the projector
\eq \mathbb{P}=\sum_{j=\mu}^{N+\mu}\sum_{m=-j}^j\ketbra{j,m} \qe
 when acting with these operators to ensure that we stay inside $\mathcal{F}_{\mu,N}$.}
From the previous section we know that all rank $1$ Hermitian tensor operators $\hat{X}_i$, $[\hat{J}_i,\,\hat{X}_j]=i\epsilon_{ijk} \hat{X}_k$ can be written as
\begin{align}
\hat{X}_3 &=a^1 a^2 \hat{C}(\hat{N})+\hat{A}(\hat{N})\hat{J}_3+\hat{C}^\dagger(\hat{N}) \ad_1\ad_2\non \\
\hat{X}_- &=-a^1 a^1 \hat{C}(\hat{N})+\hat{A}(\hat{N})\hat{J}_- +\hat{C}^\dagger(\hat{N}) \ad_2\ad_2\label{genXform} \\
\hat{X}_+ &=a^2 a^2 \hat{C}(\hat{N})+\hat{A}(\hat{N})\hat{J}_+ -\hat{C}^\dagger(\hat{N}) \ad_1\ad_1\non~,
\end{align}
for two functions of the number operator $\hat{N}=\ad_\alpha a^\alpha$, $\hat{C}(\hat{N})$ and $\hat{A}(\hat{N})$, related to the reduced matrix elements of the previous section. The $m$ dependence of the Clebsch-Gordan coefficients is precisely that of equations (\ref{osc1}) to (\ref{osc7}). Different choices of the operators $\hat{C}(\hat{N})$ and $\hat{A}(\hat{N})$ lead to different commutators of the coordinate operators $\hat{X}_i$.
For the fuzzy sphere a common choice is \eq\hat{C}(\hat{N})= 0~,\qquad \hat{A}(\hat{N})=\frac{2}{\sqrt{\hat{N}(\hat{N}+2)}}\qe
and the algebra generated, $\hat{\mathcal{A}}$, (acting on $\mathcal{F}$) is a direct sum of fuzzy sphere algebras of different cut-offs.

Such quadratics terms are used for example to form the ($Sp(6,\mathbb{R})$ component of the) dynamical group of the harmonic oscillator \cite{Wybourne} but in this case the coefficients are trivial. In the next section, will find the functions $\hat{C}(\hat{N})$ and $\hat{A}(\hat{N})$ that give rise to the commutations relations of $so(4)$, $so(3,1)$ and $e(3)$. The case of $so(4)$ has of course a simpler description in terms of two sets of Schwinger oscillators $a^\al,~\ad_\al$ and $b^\al,~b^\dagger_\al$, $\al=1,2,$ due to the isomorphism $so(4)=su(2)\oplus su(2)$. Here, however, the realization is more general and can be applied to $so(3,1)$ and $e(3)$. See \cite{Dolan:2006tx} for another case, where a number operator dependent coefficient is necessary to obtain the appropriate commutation relations. In that case the Heisenberg algebra is satisfied by composite oscillators made up of the antisymmetric product of several creation or annihilation operators and a non-trivial coefficient.

\section{The Snyder algebra}
\label{snyder}
In this section we will review the (Euclidean) Snyder algebra\footnote{In Snyder's original construction, time is included and the noncommuting coordinates are covariant under Lorentz transformation. He also extends this algebra by adding commuting momentum operators.} \cite{Snyder:1946qz}
\begin{align}\label{synderNCalgebra}
[\hat{J}_i ,\, \hat{J}_j]&=i\epsilon_{ijk}\hat{J}_k~,\\
[\hat{J}_i ,\, \hat{X}_j]&=i\epsilon_{ijk}\hat{X}_k~,\\
[\hat{X}_i ,\, \hat{X}_j]&=i\theta\epsilon_{ijk}\hat{J}_k
\end{align}
and apply the previous realization to it.
Here, $\theta\in\mathbb{R}$ is the deformation parameter with dimensions of length squared.

We can determine the operators $\hat{A}(\hat{N})$ and $\hat{C}(\hat{N})$ by comparing the last of these commutators with the commutator satisfied by the operators $\hat{X}_\pm$ of equations (\ref{genXform}) above (the remaining commutators are easily given by the action of $\hat{J}_i$) :
\begin{widetext}
\begin{align} [\hat{X}_- ,\, \hat{X}_+]&=2\left((\hat{N}+3)|\hat{C}(\hat{N}+2)|^2-(\hat{N}-1)|\hat{C}(\hat{N})|^2-\hat{A}(\hat{N})^2\right)\,\hat{J}_3 \label{X-+commutator}\\&+a^1a^2\,\hat{C}(\hat{N})\left((\hat{N}-2)\hat{A}(\hat{N}-2)-(\hat{N}+2)\hat{A}(\hat{N})\right)+ \left((\hat{N}-2)\hat{A}(\hat{N}-2)-(\hat{N}+2)\hat{A}(\hat{N})\right)\hat{C}^\dagger(\hat{N}) \,\ad_1\ad_2~.\non\end{align}

We also present the general `radius' formula
\eq
\hat{X}^2=\frac{1}{2}(\hat{N}+2)(\hat{N}+3)|\hat{C}(\hat{N}+2)|^2+\frac{1}{2}\hat{N}(\hat{N}-1)|\hat{C}(\hat{N})|^2+\frac{1}{4}\hat{N}(\hat{N}+2)\hat{A}(\hat{N})^2
\qe
\end{widetext}
and the contraction with the angular momentum operators,
\eq \hat{J}\cdot\hat{X}=\hat{X}\cdot\hat{J}=\frac{1}{4}\hat{N}(\hat{N}+2)\hat{A}(\hat{N})~. \qe

\comment{We also have the relations satisfied by the operators $\hat{Y}_i$
\begin{align}\label{Ycoordrealations}
[\hat{Y}_i,\,\hat{Y}_j]&=i\epsilon_{ijk}\left( ((\hat{N}-1)|\hat{C}(\hat{N})|^2-(\hat{N}+3)|\hat{C}(\hat{N}+2)|^2\right)\hat{J}_k\\
\hat{Y}^2&=\frac{1}{2}(\hat{N}+2)(\hat{N}+3)|\hat{C}(\hat{N}+2)|^2+\frac{1}{2}\hat{N}(\hat{N}-1)|\hat{C}(\hat{N})|^2\\
\hat{J}\cdot\hat{Y}&=\hat{Y}\cdot\hat{J}=0~.
\end{align}
}

\comment{
For example, if we want to think of the operators $\hat{X}_i$ as unit coordinates, we can impose the condition, $\hat{X}^2=1$ and say, $\hat{A}(\hat{N})=0$. If the lowest value of angular momentum is $\mu$, we find that we must choose
\eq |\hat{C}(\hat{N})|^2=\frac{\hat{N}-2\mu(-1)^{\frac{\hat{N}-2\mu}{2}}}{(\hat{N}-1)\hat{N}(\hat{N}+1)}\qe
in which case
\begin{align}
[\hat{X}_i,\,\hat{X}_j]&=i\epsilon_{ijk}\frac{4\mu(-1)^{\frac{\hat{N}-2\mu}{2}+1}}{\hat{N}(\hat{N}+2)}\hat{J}_k\\
\hat{X}^2&=1 \quad \mathrm{on}~\mathcal{F}_{\mu,\infty}.
\end{align}
Observe that $\hat{U}\hat{X}_i\hat{U}^\dagger$ satisfy similar relations to (\ref{Ycoordrealations}) (with $\hat{C}(\hat{N})$ replaced by $\hat{U}(\hat{N}-2)\hat{C}(\hat{N})\hat{U}^\dagger(\hat{N})$) and can be associated with non unit coordinate operators.

Let us next examine some choices which give rise to more beautiful commutation relations.
}

\subsection*{The Euclidean algebra}
Let us first consider the case $\theta=0$. Then the operators $\hat{X}_i$ commute and we arrive at an infinite dimensional representation of the algebra of the Euclidean group $E(3)$ (formally identical to the group of rotations and translations)\footnote{For a discusion of the Euclidean group in the context of magnetic charge quantization, see \cite{Lipkin:1969ck,Peshkin:1971bg}.}
\begin{align}
 [\hat{X}_i,\,\hat{X}_j]&=0~,\\
[\hat{J}_i,\,\hat{X}_j]&=i\epsilon_{ijk}\hat{X}_k~,\\
[\hat{J}_i,\,\hat{J}_j]&=i\epsilon_{ijk}\hat{J}_k~.
\end{align}
Representations of this algebra are labelled by the eigenvalues of the Casimirs $\hat{X}^2$ and $\hat{X}\cdot\hat{J}$, while the Hilbert space carrying the representation can be labelled by the eigenvalues of the mutually commuting operators
$\hat{X}^2,\,\hat{X}\cdot\hat{J},\,\hat{J}^2,\,\hat{J}_3$~.

Let us choose the operators $\hat{C}(\hat{N})$ and $\hat{A}(\hat{N})$ such that the coordinate operators (\ref{genXform}) commute over $\mathcal{F}$. We find
\eq \hat{A}(\hat{N})=0~, \qquad \hat{C}(\hat{N})=\frac{r e^{i\hat{\theta}({\hat{N}})}}{\sqrt{(\hat{N}-1)(\hat{N}+1)}}~,\qe
with $0<r\in\mathbbm{R}$~.
A basis for $\mathcal{F}$ is then given by $\ket{r,j,m}$. The Casimirs $\hat{X}^2$ and $\hat{X}\cdot\hat{J}$ have eigenvalues $r^2$ and $0$ respectively. Clearly, we do not have access to all the representations.

Let us raise the lower angular momentum lower bound from $0$ to $j_0=|\mu|$, with $2\mu\in \mathbbm{Z}$ as before and we denote the Hilbert space $\mathcal{F}_{j_0,\infty}$. We must therefore impose the additional condition $\hat{C}(\hat{N})\ket{j_0,m}=0$ so that $\mathcal{F}_{j_0,\infty}$ is closed under the action of the coordinate operators.
Examining the commutator (\ref{X-+commutator}),
the only possible choice is found to be
\begin{align} \hat{A}(\hat{N})&=\frac{4\mu r}{\hat{N}(\hat{N}+2)}~, \\ \hat{C}(\hat{N})&=r e^{i\hat{\theta}({\hat{N}})}\sqrt{\frac{\hat{N}^2-4\mu^2}{(\hat{N}-1)\hat{N}^2(\hat{N}+1)}}~,\end{align}
for which we find the Casimir eigenvalues
\begin{align} \hat{X}^2\ket{\mu,r,j,m}&=r^2\ket{\mu,r,j,m},\\ \hat{X}\cdot\hat{J}\ket{\mu,r,j,m}&=\mu r\ket{\mu,r,j,m}~.\end{align}
Thus we have a realization of the all the infinite dimensional representations of $e(3)$.
We may think of the eigenvalue $r$ as the classical radius, as it the eigenvalue of $\hat{X}^2$ when $\theta=0$.
The Hilbert space basis can be normalised so that
\eq \braket{\mu',r',j',m'}{\mu,r,j,m}=\delta(r-r')\delta_{\mu\mu'}\delta_{jj'}\delta_{mm'}~,\qe
while the phase $e^{i\hat{\theta}({\hat{N}})}$ is arbitrary and can be absorbed by the creation and annihilation operators. The corresponding completeness relation is
\eq \mathbbm{1}=\sum_{\mu,j,m}\int_0^\infty\d r\ketbra{r,\mu,j,m}~.\qe

\comment{
If now re-instate the angular momentum cut-off, $j_{1}$, we find that the algebraic relations become
\begin{align}
[\hat{X}_i ,\, \hat{X}_j]\ket{\mu,r,j,m}&=i\epsilon_{ijk}\delta(\hat{N}-2j_{1})r^2\frac{(j_1+1)^2-\mu^2}{(2j_1+1)(j_1+1)^2}\hat{J}_k\ket{\mu,r,j,m}~,\\
\hat{X}^2\ket{\mu,r,j,m}&=\left[r^2-\frac{r^2}{2}\delta(\hat{N}-2j_{1})\frac{(j_1+1)^2-\mu^2}{(2j_1+1)(j_1+1)^2}\right]\ket{\mu,r,j,m}~.
\end{align}
}

\subsection*{The special orthogonal algebras $so(3,1)$ and $so(4)$}
If $\theta<0$, then the Snyder algebra is isomorphic to the non-compact algebra $so(3,1)$. We again attempt to choose the operators $\hat{C}({\hat{N}})$ and $\hat{A}({\hat{N}})$ in order to obtain a representation of the algebra over $\mathcal{F}_{j_0,\infty}$.
An infinite-dimensional (bounded below) representation for $so(3,1)$ is specified by choosing
\begin{align} \hat{A}(\hat{N})&=\frac{4\mu r}{\hat{N}(\hat{N}+2)}~,\\ \hat{C}(\hat{N})&=e^{i\hat{\theta}({\hat{N}})}\sqrt{\frac{(4r^2-\theta\hat{N}^2)(\hat{N}^2-4\mu^2)}{4(\hat{N}-1)\hat{N}^2(\hat{N}+1)}}~,\end{align}
which gives
\begin{align}
\hat{X}^2&=-\frac{\theta}{4}\hat{N}(\hat{N}+2)+\theta\mu^2+r^2-\theta\\
\hat{X}\cdot\hat{J}&=\mu r~.
\end{align}
Here, like for $e(3)$, $r$ is a (non-compact) real non-negative parameter, while $j_0=|\mu|$ is a whole or half integer. Thus, the noncommutative radius $\hat{X}^2$ increases with angular momentum. Note also that for a given representation (fixed values or $\mu$ and $r$) its spectrum is discrete and that if $\mu r=0$ then the usual condition $\hat{J}\cdot\bm{x}=\bm{x}\cdot\hat{J}=0$ is satisfied.

Recalling that $\theta$ is the noncommutative parameter, it is tempting to think of the operator $\hat{X}^2$ as a $\theta$ and angular momentum dependent modification to the classical radius-squared, $r^2$.
For fixed eigenvalue of the noncommutative radius, $x^2$ and negative $\theta$, there are finitely many associated classical radii and angular momenta
\eq \left(r^2,j\right): \left(x^2+\theta(|\mu|+1),\,|\mu|\right),\,\ldots \left(r_{min}^2,\,j_{max} \right)~,\qe
where $j_{max}$ is the largest value of $j$ of such that $x^2-\theta(\mu^2-1)+\theta j(j+1)\geq 0$ and $r_{min}^2=x^2-\theta(\mu^2-1)+\theta j_{max}(j_{max}+1)$. The number of states associated to the noncommutative radius is an increasing function. However, it increases in steps due to the previous inequality. The situation is reversed for positive $\theta$; for fixed classical radius $r$, there are finitely many associated noncommutative radius values.

However, we must be more careful when $\theta$ is positive, as \eq|\hat{C}(\hat{N})|^2=\frac{(4r^2-\theta\hat{N}^2)(\hat{N}^2-4\mu^2)}{4(\hat{N}-1)\hat{N}^2(\hat{N}+1)}\qe can become negative. We require a maximum value of $j$ (recall that the eigenvalue of $\hat{N}$ is $2j$),  ${j_{1}}$. Furthermore, in order to remain inside the subspace $\mathcal{F}_{j_0,j_{1}}$, the condition $\hat{C}(\hat{N})\ket{\mu,r,j_{1}+1,m}=0$ must be imposed. We therefore find that $r$ must take the value $\sqrt{\theta}(j_{1}+1)$~. Representations and hence labelled by two discrete parameters $\mu$ and $j_1$. We have
\begin{align} \hat{A}(\hat{N})&=\frac{4\mu\sqrt{\theta}(j_{1}+1)}{\hat{N}(\hat{N}+2)}~, \\ \hat{C}(\hat{N})&= e^{i\hat{\theta}({\hat{N}})}\sqrt{\frac{\theta(4(j_{1}+1)^2-\hat{N}^2)(\hat{N}^2-4\mu^2)}{4(\hat{N}-1)\hat{N}^2(\hat{N}+1)}}~.\end{align}
Thus, we have a finite dimensional representation of dimension $j_{1}-|\mu|+1$~. This is not surprising since, when $\theta>0$, the algebra (\ref{synderNCalgebra}) is isomorphic to $so(4)=su(2)\oplus su(2)$.
The basis for these two commuting $su(2)$ algebras are given by $\hat{L}_i=(\hat{J}_i+\frac{1}{\sqrt{\theta}}\hat{X}_i)/2$ and $\hat{L}'_i=(\hat{J}_i-\frac{1}{\sqrt{\theta}}\hat{X}_i)/2$.

Note that in contrast to the standard $su(2)\oplus su(2)$ picture, where two sets of oscillators $a^\al,~\ad_\al$ and $b^\al,~b^\dagger_\al$, $\al=1,2$ are used and are associated with each copy of $su(2)$, $\hat{L}_i=\ad_\alpha (\frac{\sigma_i}{2})^\alpha{}_\beta a^\beta$  $\hat{L'}_i=b^\dagger_\alpha (\frac{\sigma_i}{2})^\alpha{}_\beta b^\beta$, here we use only one set and have terms quadratic in both creation and annihilation operators. The matrix elements in this $so(4)$ formalism are well known however \cite{Adams19881}\nocite{Bohm1988}.

We label the Hilbert space by $\ket{\mu,j_1,j,m}$ and find that
\begin{align}
\hat{X}^2&=\left(-\frac{\theta}{4}\hat{N}(\hat{N}+2)+\theta\mu^2+\theta(j_{1}+1)^2-\theta\right)\\
\hat{X}\cdot\hat{J}&=\mu\sqrt{\theta}(j_{1}+1)
\end{align}
The two Casimirs are
\begin{align}
\hat{L}^2 &=\frac{1}{4}(\mu+j_{1})(\mu+j_{1}+2)\\
\hat{L'}^2&=\frac{1}{4}(\mu-j_{1})(\mu-j_{1}-2)
\end{align}
Observe that in this case the spectrum of $\hat{X}^2$ decreases with $j$. Hence, higher angular momentum is associated with a smaller noncommutative radius.
To recover the continuum limit and obtain a infinite dimensional representation of the Euclidean algebra, we take $\theta\rightarrow 0$ and $j_{1}\rightarrow \infty$ in such a way that $\sqrt{\theta}(j_{1}+1)$ remains a constant. Hence $r=\sqrt{\theta}(j_{1}+1)$ is again referred to as the classical radius.

The question of the Laplacian might now be considered. However, we require an analogue of the continuum radial derivative and to this end, in the next section, we shall modifying the Snyder algebra so that it can be extended with the generator of dilations. In the continuum, this generator is composed of the radius and its derivative.

\section{Generalising Snyder's algebra}
\label{gensny}
Let us now return to the modification the Snyder algebra introduced in (\ref{JJcommutator}), (\ref{JXcommutator}) and (\ref{genXXcommutator}) :
\begin{align}\label{gensnyderalg}
[\hat{J}_i ,\, \hat{J}_j]&=i\epsilon_{ijk}\hat{J}_k~,\\
[\hat{J}_i ,\, \hat{X}_j]&=i\epsilon_{ijk}\hat{X}_k~,\\
[\hat{X}_i ,\, \hat{X}_j]&=i\epsilon_{ijk}(\alpha\hat{J}_k+\beta\hat{X}_k)~.
\end{align}
It is easily seen that the algebra (\ref{gensnyderalg}) is isomorphic to the Snyder algebra (\ref{synderNCalgebra}) and so it is isomorphic to
\eq \begin{array}{ll}
so(4) & \mathrm{for}\quad\alpha+\frac{\beta^2}{4}>0\\
e(3) & \mathrm{for}\quad\alpha+\frac{\beta^2}{4}=0\\
so(3,1) & \mathrm{for}\quad\alpha+\frac{\beta^2}{4}<0~~.
\end{array}\qe
Once again we can find the necessary form for the operators $\hat{A}(\hat{N})$ and $\hat{C}(\hat{N})$ to be
\begin{align} \hat{A}(\hat{N})&=\frac{\beta \hat{N}(\hat{N}+2)+8\mu r}{2\hat{N}(\hat{N}+2)}~, \\ |\hat{C}(\hat{N})|^2&=\frac{((-\alpha-\beta^2 /4)\hat{N}^2+4 r^2)(\hat{N}^2-4\mu^2)}{4(\hat{N}-1)\hat{N}^2 (\hat{N}+1)}~,\end{align}
while the radius is
\eq \hat{X}^2=-\frac{\alpha}{4}\hat{N}(\hat{N}+2)-(\alpha+\beta^2/4)(1-\mu^2)+\beta\mu r+r^2~. \qe
The representations are labelled by $r\geq0$ and $2\mu\in\mathbb{Z}$.

The special case $\alpha=-\beta^2/4$ is worth special study. The algebra is then isomorphic to that of $e(3)$. The formulae for the radius and $\hat{C}(\hat{N})$ simplify and in particular $\hat{C}(\hat{N})$ no longer depends on the noncommutative parameter and contains an overall factor of $r$. The coordinate operators $\hat{X}_i$ then satisfy the property
\eq r\dd{}{r} (\hat{X}_i)=\hat{X}_i-\frac{\beta}{2}\hat{J}_i~. \label{rderivative}\qe
The derivative with respect to the classical radius is an automorphism of the algebra.
This hints at some further structure, which we will now elucidate.

The Laplacian in three dimensions can be written in terms of the generators of translations,
\eq \Delta=\sum_{i=1}^3\dd{{}^2}{x_i^2}~,\qe or alternatively, in terms of the radial coordinate and the generators of rotations and homogeneous dilations (uniform/isotropic scalings),
\begin{align}
\Delta &=\left(\frac{1}{r}\dd{}{r}r\right)^2-\frac{1}{r^2}\hat{J}^2\\
&=-\frac{1}{r}\left(\hat{D}^2+\frac{1}{4}\right)\frac{1}{r}-\frac{1}{r^2}\hat{J}^2~,\label{laplacian-dilations}
\end{align}
where $\hat{D}=-i(r\dd{}{r}+\frac{3}{2})$ is the Hermitian generator of dilations and $\hat{J}_i$ generate rotations. 
All these objects are contained in the classical similarity group of translations, rotations and homogeneous dilations, which has the algebra
\begin{align}
[\hat{J}_i ,\, \hat{J}_j]&=i\epsilon_{ijk}\hat{J}_k~,\qquad[\hat{J}_i ,\, \hat{p}_j]=i\epsilon_{ijk}\hat{p}_k~,
\\
[\hat{p}_i ,\, \hat{p}_j]&=0~,\qquad[\hat{D},\, \hat{J}_i]=0~,\qquad [\hat{D} ,\, \hat{p}_i]=-i\hat{p}_i~.\non
\end{align}
Replacing the generators of translations, $\hat{p}_i$, with the coordinates themselves, leads to a formally identical algebra.

We will then consider an enlargement of the algebra (\ref{gensnyderalg}) by adding the, generator of dilations $\hat{D}$, specified only via the commutators
\eq [\hat{D} ,\, \hat{J}_i]=0~, \qquad [\hat{D} ,\, \hat{X}_i]=-i(\hat{X}_i+\gamma \hat{J}_i)~.\qe
We have chosen to maintain our ansatz of spherical symmetry and have only modified the space commutator. Application of the Jacobi identity provides constraints on the parameters $\alpha$, $\beta$ and $\gamma$
\eq \gamma=-\frac{\beta}{2}~,\qquad \alpha=\frac{\beta\gamma}{2}~,\qe
which implies precisely the special case $\alpha=-\beta^2/4$.

Thus, an attempt to combine dilations with the Snyder algebra via the previous commutators, leads one to modify the coordinate-coordinate commutator as in (\ref{gensnyderalg}) with $\alpha=-\beta^2/4$, so that the coordinate and rotation operators satisfy an algebra isomorphic to $e(3)$. Including dilations, the algebra is formally identical to the classical similarity algebra
\begin{align}\label{NCsimilarity}
[\hat{X}_i ,\, \hat{X}_j]&=i\epsilon_{ijk}(-\frac{\beta^2}{4}\hat{J}_k+\beta\hat{X}_k)~,\\
[\hat{J}_i ,\, \hat{J}_j]&=i\epsilon_{ijk}\hat{J}_k~,\qquad [\hat{J}_i ~,\, \hat{X}_j]=i\epsilon_{ijk}\hat{X}_k~,\\
[\hat{D},\, \hat{J}_i]&=0~,\qquad [\hat{D} ~,\, \hat{X}_i]=-i(\hat{X}_i-\frac{\beta}{2} \hat{J}_i)~.
\end{align}
We note that this is simply a transformation of the classical similarity algebra\footnote{Since the algebra is isomorphic to the standard similarity algebra, we can introduce momentum operators $\hat{p}_i$ and write the coordinate operators in term of oscillators $a^i=\frac{1}{\sqrt{2}}(\hat{x}_i+i\hat{p}_i)$, $\ad_i=\frac{1}{\sqrt{2}}(\hat{x}_i-i\hat{p}_i)$, $i=1,2,3$. The angular momentum operators are associated to the $su(2)$ subalgebra, $\hat{J}_i=-i\epsilon_{ijk}\ad_ja^k$, and the Laplacian $\hat{p}^2$ is invariant under the corresponding subgroup. The realisation we have presented above however, only describes that part of the Hilbert space generated by the (noncommuting) coordinates.} (taking the classical coordinate $\hat{x}_i$ as a generator instead of $\hat{p}_i$): $\hat{X}_i =\hat{x}_i+\frac{\beta}{2}\hat{J}_i$ \footnote{If one attempts to add commuting momentum operators $\hat{P}_i$ to the Snyder algebra, then there are infinitely many ways to choose the $[\hat{X}_i,\,\hat{P}_j]$ and still have a consistent algebra \cite{Battisti:2008xy}. Here, we have modified the Snyder algebra such that it is isomorphic to $e(3)$. We then have a natural choice for this commutator: $[\hat{X}_i,\,\hat{P}_j]=i\delta_{ij}+i\frac{\beta}{2}\epsilon_{ijk}\hat{P}_k$.}.
Furthermore, comparing the last commutator above with equation (\ref{rderivative}), we see that our realization of the algebra (\ref{gensnyderalg}) can be extended to the algebra (\ref{NCsimilarity}) by taking
\eq \hat{D}=-i(r\dd{}{r}+\frac{3}{2})~,\label{dilations}\qe
along with the coordinate operators
\begin{align}
\hat{X}_3 &=a^1 a^2 \hat{C}(\hat{N})+\hat{A}(\hat{N})(\ad_1 a^1 -\ad_2 a^2)/2 +\hat{C}^\dagger(\hat{N}) \ad_1\ad_2 \non\\
\hat{X}_- &=-a^1 a^1 \hat{C}(\hat{N})+\hat{A}(\hat{N})\ad_2 a^1 +\hat{C}^\dagger(\hat{N}) \ad_2\ad_2 \label{finalrealization}\\
\hat{X}_+ &=a^2 a^2 \hat{C}(\hat{N})+\hat{A}(\hat{N})\ad_1 a^2 -\hat{C}^\dagger(\hat{N}) \ad_1\ad_1~,\non
\end{align}
where
\begin{align} \hat{A}(\hat{N})&=\frac{\beta \hat{N}(\hat{N}+2)+8\mu r}{2\hat{N}(\hat{N}+2)}~,\\ \hat{C}(\hat{N})&=r\sqrt{\frac{\hat{N}^2-4\mu^2}{(\hat{N}-1)\hat{N}^2 (\hat{N}+1)}}~.\end{align}

\comment{The dilation operator is
\eq \hat{D}=\frac{1}{2}(r\hat{p}_r+\hat{p}_r r)
\qe
where \eq [r,\,\hat{p}_r]=i \qe and $\hat{J}_i$ are the usual angular momentum operators. The operator\footnote{This operator is Hermitian on the space of radial functions satisfying $\int_0^\infty\dd{}{r}( r^2 f^*(r)g(r))\d r=\left[ r^2 f^*(r)g(r)\right]^\infty_{0} =0$. Care is needed on the issue of radial translations and negative radii. See \cite{Liboff:1976, Mosley:2003} for further details. Since we do not include $\hat{p}_r$ in the algebra but only $\hat{D}$ and $\hat{r}$, we do not need to worry about this problem here.} $\hat{p}_r$ has the representation $\hat{p}_r=-i\frac{1}{r}\dd{}{r}r$.}

The noncommutative radius is
\eq \hat{X}^2=\frac{\beta^2}{4}\hat{J}^2+\beta\mu r+r^2~,\label{radius} \qe
while
\eq \hat{X}\cdot\hat{J}=\frac{\beta}{2}\hat{J}^2+\mu r~.\qe
We also find that
\begin{align} [\hat{D},\,\hat{X}^2]&=-i(2r^2+{\beta}\mu r)=-2i(\hat{X}^2-\frac{\beta}{2}\hat{X}\cdot\hat{J})\,,\\
[\hat{D},\,\hat{X}\cdot\hat{J}]&=-i\mu r=-i(\hat{X}\cdot\hat{J}-\frac{\beta}{2}\hat{J}^2)~.\end{align}
We can consider the reducible representation of $e(3)$ obtained by allowing $r$ to take all values from zero to infinity and in this way think of the coordinates as describing a particle on a noncommutative space. The noncommutative radius is related to the classical radius by (\ref{radius}).

\section{The algebra $\hat{\mathcal{A}}$ and the Laplacian}
As mentioned briefly in section {\bf \ref{tensorop}}, the algebra, $\hat{\mathcal{A}}$, generated by the coordinate operators can be decomposed into
subspaces of the angular Laplacian
\eq \hat{M}=\sum_{l=0}^\infty \sum_{m=-l}^l\sum_d  c_{d,l,m}\hat{\Psi}^{d}_{lm}~.\qe
The tensors $\hat{\Psi}^{d}_{lm}$ are the analogue of the (smooth) continuum functions $R(r)Y_{lm}(\theta,\phi)$ where $R(r)$ is any admissible radial function and $Y_{lm}(\theta,\phi)$ are the spherical harmonics. The continuum eigenfunctions with eigenvalue $-k^2$ are written $\psi_{k,l,m}(r,\theta,\phi)=j_l(kr)Y_{lm}(\theta,\phi)=\frac{j_l(kr)}{r^l}R_{lm}(r,\theta,\phi)$ where $j_l(kr)$ are the spherical Bessel functions  and $R_{lm}(r,\theta,\phi)$ are the solid harmonics\footnote{Note that $\frac{j_l(kr)}{r^l}$ is expressable as a power series that contains only even (non-negative) powers of $r$.}.

The solid harmonics are the solutions to the continuum Laplace equation and are, up to normalization factors, just the traceless symmetric products of the coordinates. In the noncommutative case we are considering, we denote such products $\hat{R}_{lm}(\hat{X})$ and call them solid polarization tensors. Up to a normalization, $n_l$, they are written
\eq \hat{R}_{lm}(\hat{X})\equiv n_l\mathcal{P}^{j_1\cdots j_l}_{i_1\cdots i_l}\hat{X}_{j_1}\cdots\hat{X}_{j_l}~,\qe
where $\mathcal{P}^{j_1\cdots j_l}_{i_1\cdots i_l}$ is the projector onto symmetric traceless tensors. The index $m$ on the left hand side of this equation denotes the eigenvalue of $\mathcal{J}_3$ (defined below), under which the right hand side can be split up.
Due to algebra (\ref{NCsimilarity}), the algebra of the coordinate operators also contains polynomials in the operators $\hat{Y}_i=-\frac{\beta^2}{4}\hat{J}_i+\beta\hat{X}_i$. Hence, the algebra is spanned by symmetric polynomials in $\hat{X}_i$, $\hat{Y}_i$ and $\epsilon_{ijk}\hat{X}_j \hat{Y}_k$ and basis elements are specified by some normal ordering prescription. We shall not discuss these operators in more detail here, postponing it to future work. However, we will introduce a Laplacian on this algebra, discuss the zero modes and show that the correct continuum spectrum is obtained in the $\beta \longrightarrow 0$ limit. Clearly, we need only then consider the solid polarization tensors $\hat{R}_{lm}(\hat{X})$, since the other polynomials vanish in the limit. The arbitrary radial dependence comes from operator coefficients $\hat{R}(\hat{X}^2,\hat{Y}^2,\hat{X}\cdot\hat{Y})$, dependent on the scalar invariants (only $\hat{X}^2\longrightarrow r^2$ is non-vanishing in the limit).

Let us examine the possible terms that might make up the rotationally invariant noncommutative Laplacian.
We have two basic vector operators at our disposal: $\hat{X}_i$ and $\hat{J}_i$.
From them, we can form the derivations on the algebra of the coordinate operators, $\hat{\mathcal{A}}$ :
\begin{align}
\hat{\mathcal{X}}_i &=\hat{X}_i^\mathrm{L}-\hat{X}_i^\mathrm{R}\\
\hat{\mathcal{J}}_i &=\hat{J}_i^\mathrm{L}-\hat{J}_i^\mathrm{R}~,
\end{align}
where the superscripts indicate left and right actions. These operators satisfy the Leibniz condition and vanish on the identity. The operator $\hat{\mathcal{J}}_i $ is used to write the Laplacian on the fuzzy sphere, $\hat{\Delta}_{S^2_\mathrm{F}}=-\hat{\mathcal{J}}^2$.
So we are led to consider the following operators $\hat{\mathcal{J}}^2$, $\hat{\mathcal{X}}^2$ and $\hat{\mathcal{J}}\cdot\hat{\mathcal{X}}$. 
In that absence of an appropriate noncommutative analogue and motivated by (\ref{laplacian-dilations}) and (\ref{dilations}) we will also include the possibility of $r$ and its derivatives appearing in the Laplacian.

We take the following ansatz for the noncommutative Laplacian, $\hat{\Delta}$ :
\eq \hat{\Delta}=a\dd{{}^2}{r^2}+e\frac{2}{r}\dd{}{r}+\frac{b}{\beta^2r^2}\hat{\mathcal{X}}^2+\frac{c}{r^2}\hat{\mathcal{J}}^2+\frac{d}{\beta r^2}\hat{\mathcal{J}}\cdot\hat{\mathcal{X}} ~\qe
 and choose $a,b,c,d$ and $e$ such that the continuum spectrum and eigenfunctions are obtained in the $\beta \rightarrow 0$ limit. We have included a factor of $\frac{1}{r^2}$ in the `angular' part to match the continuum and corrected the dimensions with factors of $\beta$. We impose
\eq\hat{\Delta}(\hat{X}^2)^n \xrightarrow[\beta \rightarrow 0]{} 2n(2n+1)r^{2n-2}=\left(\dd{{}^2}{r^2}+\frac{2}{r}\dd{}{r}\right)r^{2n}~,\label{radiallimit}\qe
\eq\hat{\Delta}(\hat{X}_i)=0~,\qquad \hat{\Delta}(\hat{X}_i\hat{X}_j+\hat{X}_j\hat{X}_i-\frac{2}{3}\delta_{ij}\hat{X}^2)=0~.\qe

The first condition imposes $a=1$ and $4e-b=4$. Asking the coordinate operators to have eigenvalue $0$ as in the continuum, gives $b+c+d+1=0$ and $b+d+2e=0$. One further condition is necessary and we choose to also impose that the traceless symmetric products of two coordinate operators also have vanishing eigenvalue, which results in only one extra linearly independent equation, $b=-4$.
We find that
\eq a=1,\qquad b=-4,\qquad c=-1, \qquad d=4, \qquad e=0~, \qe
so we have
\begin{align} \hat{\Delta}&=\dd{{}^2}{r^2}-\frac{4}{\beta^2r^2}\hat{\mathcal{X}}^2-\frac{1}{r^2}\hat{\mathcal{J}}^2+\frac{4}{\beta r^2}\hat{\mathcal{J}}\cdot\hat{\mathcal{X}}\non\\
 &=\dd{{}^2}{r^2}-\frac{4}{\beta^2r^2}\left(\hat{\mathcal{X}}_i-\frac{\beta}{2}\hat{\mathcal{J}}_i\right)^2.\label{laplacian}\end{align}

As already mentioned, the solutions to the continuum Laplace equation are the solid harmonics, $\hat{R}_{lm}(\hat{X})$. We have already imposed $\hat{\Delta}\hat{R}_{1m}(\hat{X})=\hat{\Delta}\hat{R}_{2m}(\hat{X})=0$. From equations (\ref{symrderiv}) and (\ref{symL2comm}) in the appendix, we see that in fact all the solid polarization tensors are solutions of the noncommutative Laplace equation
\eq \hat{\Delta}\hat{R}_{lm}(\hat{X})=0~.\label{solidharmonics}\qe

We will not discuss the other eigenoperators of the Laplacian $\hat{\Delta}$ here, postponing it for future work. However, we give its value on the first few powers of the noncommutative radius and the general result up to order $\beta^2$:
\eq \hat{\Delta}(\hat{X^2})=6,\qquad \hat{\Delta}(\hat{X}^4)=20\hat{X}^2+2\beta^2\qe
\eq \hat{\Delta}(\hat{X}^6)=42\hat{X}^4+14\beta^2 \hat{X}^2-8\beta^2 r^2-8\beta^3 \mu r+\beta^4 (1-\mu^2)~. \qe
\begin{align} \hat{\Delta}(\hat{X}^{2n})&=2n(2n+1)\hat{X}^{2n-2}+\frac{\beta^2}{3}n(n-1)\hat{X}^{2n-8}\times\non\\&\left(\vphantom{\hat{X}^4}(n^2-n+1)\hat{X}^4-2(n-1)(n-2)r^2\hat{X}^2\right.\non\\&\left.+(n-2)(n-3)r^4\vphantom{\hat{X}^4}\right)+\mathcal{O}(\beta^3)~. \label{truncatedradiusspectrum}\end{align}
The full expansion being derivable from equations (\ref{radius}), (\ref{radialderivative}) and (\ref{J2ncomm}).
Notice the noncommutative corrections to the classical result.
\comment{Due to the commutation relations satisfied by the coordinate operators, the algebra they generate, $\hat{\mathcal{A}}$, is spanned by polynomials in both the operators $\hat{X}_i$ and $\beta\hat{X}_i-\frac{\beta^2}{4}\hat{J}_i$. Hence, there are eigenoperators of the Laplacian which are purely noncommutative, i.e.{ }which vanish in the continuum limit, and only these eigenoperators contain factors of the angular momentum $\hat{J}_i$.}
Equations (\ref{solidharmonics}) and (\ref{truncatedradiusspectrum}) show that the correct continuum spectrum ($-k^2$) and eigenfunctions ($j_l(kr)Y_{lm}(\theta, \phi)$) are obtained in the limit.

\section{Outlook}
Further work is needed to obtain the full noncommutative spectrum of the Laplacian (\ref{laplacian}). As can be seen from equations (\ref{truncatedradiusspectrum}) and (\ref{J2ncomm}), the calculations are cumbersome and require the use of the realization to make progress. In order to apply a projection to the algebra in the spirit of the fuzzy disc, it may be useful to work with coherent states, which in this case would be associated with the group $E(3)$. Such coherent states have been constructed by Bi\`evre \cite{bievre:1401} and Isham and Klauder \cite{isham:607}. Applying a projection of the algebra onto a finite range of angular momentum is then expected to lead to a finite noncommutative (phase space) algebra associated to a particle on a fuzzy sphere of radius $r$ that is smeared out in the radial direction according to the angular momentum of the particle as in (\ref{radius}).

\begin{acknowledgments}
This work was supported by the Belgian Federal Office for Scientific, Technical and Cultural Affairs through the Interuniversity Attraction Pole P6/11. S.M.{} would like to thank Larissa Lorenz for fruitful discussions.
\end{acknowledgments}

\appendix*
\section{Formulae}
\begin{widetext}
In this appendix, we present the following useful formulae:
\begin{align}
\hat{\mathcal{X}}^2(\hat{X}_i)&=\frac{3}{2}\beta^2\hat{X}_i-\frac{1}{2}\beta^3\hat{J}_i~,& \quad \hat{\mathcal{J}}^2(\hat{X}_i)&=2\hat{X}_i~, & \quad \hat{\mathcal{J}}\cdot\hat{\mathcal{X}}(\hat{X}_i)&=2\beta\hat{X}_i-\frac{\beta^2}{2}\hat{J}_i~,\non\\
\hat{\mathcal{X}}^2(\hat{J}_i)&=2\beta\hat{X}_i-\frac{\beta^2}{2}\hat{J}_i~,&\quad \hat{\mathcal{J}}^2(\hat{J}_i)&=2\hat{J}_i~,& \quad \hat{\mathcal{J}}\cdot\hat{\mathcal{X}}(\hat{J}_i)&=2\hat{X}_i~,\\
\hat{\mathcal{X}}^2(\hat{J}^2)&=-4r^2~, &\quad \hat{\mathcal{J}}^2(\hat{J}^2)&=0 ~,&\quad \hat{\mathcal{J}}\cdot\hat{\mathcal{X}}(\hat{J}^2)&=0~,\non
\end{align}

\begin{align}
\dd{{}^2}{r^2}(\hat{X}^2)^n&=n(n-1)(\beta\mu+2r)^2(\hat{X}^2)^{n-2}+2n(\hat{X}^2)^{n-1}~,\label{radialderivative}\\
\dd{{}^2}{r^2}\hat{X}_{\{i_1}\cdots\hat{X}_{i_l\}}&=\frac{l(l-1)}{r^2}\hat{L}_{\{i_1}\hat{L}_{i_2}\hat{X}_{i_3}\cdots \hat{X}_{i_l\}}~,\label{symrderiv}\\
\hat{\mathcal{L}}^2\left(\hat{X}_{\{i_1}\cdots\hat{X}_{i_l\}}\right)&=\frac{\beta^2}{4}l(l-1)\left(\hat{L}_{\{i_1}\hat{L}_{i_2}\hat{X}_{i_3}\cdots\hat{X}_{i_l\}}-r^2\delta_{\{i_1 i_2}\hat{X}_{i_3}\cdots\hat{X}_{i_l\}}\right)~,\label{symL2comm}
\end{align}
where $\hat{L}_i=\hat{X}_i-\frac{\beta}{2}\hat{J}_i\;(=\hat{x}_i)$ and similarly for $\hat{\mathcal{L}}_i$. The $\{\cdots\}$ indicate symmetrization (without any factorial factors). The following result is most easily found by using the realization (\ref{finalrealization})
\begin{align} \hat{\mathcal{X}}^2(\hat{J}^{2n})&=-\frac{2\mu^2 r^2}{4^n}\left(4\hat{N}^{n-1}(\hat{N}+2)^{n-1}-\frac{2(\hat{N}+2)^{n-1}(\hat{N}+4)^n}{\hat{N}+1}-\frac{2\hat{N}^{n-1}(\hat{N}-2)^n}{\hat{N}+1}\right)\non\\
&+\frac{r^2}{4^n}\left(2\hat{N}^n(\hat{N}+2)^n-\frac{(\hat{N}+2)^{n+1}(\hat{N}+4)^n}{\hat{N}+1}-\frac{\hat{N}^{n+1}(\hat{N}-2)^n}{\hat{N}+1}\right)~.\label{J2ncomm}
\end{align}
The $\hat{N}$ dependence in the right hand side of this expression is expressable entirely in terms of non-negative powers of $\hat{J}^2=\frac{1}{4}\hat{N}(\hat{N}+2)$.
\end{widetext}

\bibliography{bibfile}

\begin{thebibliography}{45}%
\makeatletter
\providecommand \@ifxundefined [1]{%
 \@ifx{#1\undefined}
}%
\providecommand \@ifnum [1]{%
 \ifnum #1\expandafter \@firstoftwo
 \else \expandafter \@secondoftwo
 \fi
}%
\providecommand \@ifx [1]{%
 \ifx #1\expandafter \@firstoftwo
 \else \expandafter \@secondoftwo
 \fi
}%
\providecommand \natexlab [1]{#1}%
\providecommand \enquote  [1]{``#1''}%
\providecommand \bibnamefont  [1]{#1}%
\providecommand \bibfnamefont [1]{#1}%
\providecommand \citenamefont [1]{#1}%
\providecommand \href@noop [0]{\@secondoftwo}%
\providecommand \href [0]{\begingroup \@sanitize@url \@href}%
\providecommand \@href[1]{\@@startlink{#1}\@@href}%
\providecommand \@@href[1]{\endgroup#1\@@endlink}%
\providecommand \@sanitize@url [0]{\catcode `\\12\catcode `\$12\catcode
  `\&12\catcode `\#12\catcode `\^12\catcode `\_12\catcode `\%12\relax}%
\providecommand \@@startlink[1]{}%
\providecommand \@@endlink[0]{}%
\providecommand \url  [0]{\begingroup\@sanitize@url \@url }%
\providecommand \@url [1]{\endgroup\@href {#1}{\urlprefix }}%
\providecommand \urlprefix  [0]{URL }%
\providecommand \Eprint [0]{\href }%
\providecommand \doibase [0]{http://dx.doi.org/}%
\providecommand \selectlanguage [0]{\@gobble}%
\providecommand \bibinfo  [0]{\@secondoftwo}%
\providecommand \bibfield  [0]{\@secondoftwo}%
\providecommand \translation [1]{[#1]}%
\providecommand \BibitemOpen [0]{}%
\providecommand \bibitemStop [0]{}%
\providecommand \bibitemNoStop [0]{.\EOS\space}%
\providecommand \EOS [0]{\spacefactor3000\relax}%
\providecommand \BibitemShut  [1]{\csname bibitem#1\endcsname}%
\let\auto@bib@innerbib\@empty
\bibitem [{\citenamefont {Connes}(1994)}]{Connes:1994yd}%
  \BibitemOpen
  \bibfield  {author} {\bibinfo {author} {\bibfnamefont {A.}~\bibnamefont
  {Connes}},\ }\href@noop {} {\emph {\bibinfo {title} {Noncommutative
  Geometry}}}\ (\bibinfo  {publisher} {Academic Press},\ \bibinfo {year}
  {1994})\BibitemShut {NoStop}%
\bibitem [{\citenamefont {Doplicher}\ \emph {et~al.}(1994)\citenamefont
  {Doplicher}, \citenamefont {Fredenhagen},\ and\ \citenamefont
  {Roberts}}]{Doplicher:1994zv}%
  \BibitemOpen
  \bibfield  {author} {\bibinfo {author} {\bibfnamefont {S.}~\bibnamefont
  {Doplicher}}, \bibinfo {author} {\bibfnamefont {K.}~\bibnamefont
  {Fredenhagen}}, \ and\ \bibinfo {author} {\bibfnamefont {J.~E.}\ \bibnamefont
  {Roberts}},\ }\bibfield  {title} {\enquote {\bibinfo {title} {{Space-time
  quantization induced by classical gravity}},}\ }\href {\doibase
  10.1016/0370-2693(94)90940-7} {\bibfield  {journal} {\bibinfo  {journal}
  {Phys. Lett.}\ }\textbf {\bibinfo {volume} {B331}},\ \bibinfo {pages}
  {39--44} (\bibinfo {year} {1994})}\BibitemShut {NoStop}%
\bibitem [{\citenamefont {Doplicher}\ \emph {et~al.}(1995)\citenamefont
  {Doplicher}, \citenamefont {Fredenhagen},\ and\ \citenamefont
  {Roberts}}]{Doplicher:1994tu}%
  \BibitemOpen
  \bibfield  {author} {\bibinfo {author} {\bibfnamefont {Sergio}\ \bibnamefont
  {Doplicher}}, \bibinfo {author} {\bibfnamefont {Klaus}\ \bibnamefont
  {Fredenhagen}}, \ and\ \bibinfo {author} {\bibfnamefont {John~E.}\
  \bibnamefont {Roberts}},\ }\bibfield  {title} {\enquote {\bibinfo {title}
  {The quantum structure of space-time at the {P}lanck scale and quantum
  fields},}\ }\href@noop {} {\bibfield  {journal} {\bibinfo  {journal} {Commun.
  Math. Phys.}\ }\textbf {\bibinfo {volume} {172}},\ \bibinfo {pages}
  {187--220} (\bibinfo {year} {1995})},\ \Eprint
  {http://arxiv.org/abs/hep-th/0303037} {hep-th/0303037} \BibitemShut {NoStop}%
\bibitem [{\citenamefont {Douglas}\ and\ \citenamefont
  {Nekrasov}(2001)}]{Douglas:2001ba}%
  \BibitemOpen
  \bibfield  {author} {\bibinfo {author} {\bibfnamefont {Michael~R.}\
  \bibnamefont {Douglas}}\ and\ \bibinfo {author} {\bibfnamefont {Nikita~A.}\
  \bibnamefont {Nekrasov}},\ }\bibfield  {title} {\enquote {\bibinfo {title}
  {Noncommutative field theory},}\ }\href@noop {} {\bibfield  {journal}
  {\bibinfo  {journal} {Rev. Mod. Phys.}\ }\textbf {\bibinfo {volume} {73}},\
  \bibinfo {pages} {977--1029} (\bibinfo {year} {2001})},\ \Eprint
  {http://arxiv.org/abs/hep-th/0106048} {hep-th/0106048} \BibitemShut {NoStop}%
\bibitem [{\citenamefont {Seiberg}\ and\ \citenamefont
  {Witten}(1999)}]{Seiberg:1999vs}%
  \BibitemOpen
  \bibfield  {author} {\bibinfo {author} {\bibfnamefont {Nathan}\ \bibnamefont
  {Seiberg}}\ and\ \bibinfo {author} {\bibfnamefont {Edward}\ \bibnamefont
  {Witten}},\ }\bibfield  {title} {\enquote {\bibinfo {title} {String theory
  and noncommutative geometry},}\ }\href@noop {} {\bibfield  {journal}
  {\bibinfo  {journal} {JHEP}\ }\textbf {\bibinfo {volume} {09}},\ \bibinfo
  {pages} {032} (\bibinfo {year} {1999})},\ \Eprint
  {http://arxiv.org/abs/hep-th/9908142} {hep-th/9908142} \BibitemShut {NoStop}%
\bibitem [{\citenamefont {Myers}(1999)}]{Myers:1999ps}%
  \BibitemOpen
  \bibfield  {author} {\bibinfo {author} {\bibfnamefont {Robert~C.}\
  \bibnamefont {Myers}},\ }\bibfield  {title} {\enquote {\bibinfo {title}
  {Dielectric-branes},}\ }\href@noop {} {\bibfield  {journal} {\bibinfo
  {journal} {JHEP}\ }\textbf {\bibinfo {volume} {12}},\ \bibinfo {pages} {022}
  (\bibinfo {year} {1999})},\ \Eprint {http://arxiv.org/abs/hep-th/9910053}
  {hep-th/9910053} \BibitemShut {NoStop}%
\bibitem [{\citenamefont {Dunne}\ and\ \citenamefont
  {Jackiw}(1993)}]{Dunne:1992ew}%
  \BibitemOpen
  \bibfield  {author} {\bibinfo {author} {\bibfnamefont {Gerald~V.}\
  \bibnamefont {Dunne}}\ and\ \bibinfo {author} {\bibfnamefont
  {R.}~\bibnamefont {Jackiw}},\ }\bibfield  {title} {\enquote {\bibinfo {title}
  {{`Peierls substitution' and Chern-Simons quantum mechanics}},}\ }\href@noop
  {} {\bibfield  {journal} {\bibinfo  {journal} {Nucl. Phys. Proc. Suppl.}\
  }\textbf {\bibinfo {volume} {33C}},\ \bibinfo {pages} {114--118} (\bibinfo
  {year} {1993})},\ \Eprint {http://arxiv.org/abs/hep-th/9204057}
  {arXiv:hep-th/9204057} \BibitemShut {NoStop}%
\bibitem [{\citenamefont {Bander}(2004)}]{Bander:2004nj}%
  \BibitemOpen
  \bibfield  {author} {\bibinfo {author} {\bibfnamefont {Myron}\ \bibnamefont
  {Bander}},\ }\bibfield  {title} {\enquote {\bibinfo {title} {{Noncommuting
  spherical coordinates}},}\ }\href {\doibase 10.1103/PhysRevD.70.087702}
  {\bibfield  {journal} {\bibinfo  {journal} {Phys. Rev.}\ }\textbf {\bibinfo
  {volume} {D70}},\ \bibinfo {pages} {087702} (\bibinfo {year} {2004})},\
  \Eprint {http://arxiv.org/abs/hep-th/0407177} {arXiv:hep-th/0407177}
  \BibitemShut {NoStop}%
\bibitem [{\citenamefont {Frenkel}\ and\ \citenamefont
  {Pereira}(2004)}]{Frenkel:2004ff}%
  \BibitemOpen
  \bibfield  {author} {\bibinfo {author} {\bibfnamefont {J.}~\bibnamefont
  {Frenkel}}\ and\ \bibinfo {author} {\bibfnamefont {S.~H.}\ \bibnamefont
  {Pereira}},\ }\bibfield  {title} {\enquote {\bibinfo {title} {{Coordinate
  noncommutativity in strong non-uniform magnetic fields}},}\ }\href {\doibase
  10.1103/PhysRevD.69.127702} {\bibfield  {journal} {\bibinfo  {journal} {Phys.
  Rev.}\ }\textbf {\bibinfo {volume} {D69}},\ \bibinfo {pages} {127702}
  (\bibinfo {year} {2004})},\ \Eprint {http://arxiv.org/abs/hep-th/0401048}
  {arXiv:hep-th/0401048} \BibitemShut {NoStop}%
\bibitem [{\citenamefont {Govaerts}\ and\ \citenamefont
  {Murray}(2010)}]{Govaerts:2009ri}%
  \BibitemOpen
  \bibfield  {author} {\bibinfo {author} {\bibfnamefont {Jan}\ \bibnamefont
  {Govaerts}}\ and\ \bibinfo {author} {\bibfnamefont {Se\'an}\ \bibnamefont
  {Murray}},\ }\bibfield  {title} {\enquote {\bibinfo {title} {Noncommuting
  coordinates and magnetic monopoles},}\ }\href {\doibase
  10.1007/JHEP01(2010)008} {\bibfield  {journal} {\bibinfo  {journal} {JHEP}\
  }\textbf {\bibinfo {volume} {01}},\ \bibinfo {pages} {008} (\bibinfo {year}
  {2010})},\ \Eprint {http://arxiv.org/abs/0910.4356} {arXiv:0910.4356
  [hep-th]} \BibitemShut {NoStop}%
\bibitem [{\citenamefont {Balachandran}\ \emph {et~al.}(2005)\citenamefont
  {Balachandran}, \citenamefont {K\"{u}rk\c{c}\"{u}o\v{g}lu},\ and\
  \citenamefont {Vaidya}}]{Balachandran:2005ew}%
  \BibitemOpen
  \bibfield  {author} {\bibinfo {author} {\bibfnamefont {A.~P.}\ \bibnamefont
  {Balachandran}}, \bibinfo {author} {\bibfnamefont {S.}~\bibnamefont
  {K\"{u}rk\c{c}\"{u}o\v{g}lu}}, \ and\ \bibinfo {author} {\bibfnamefont
  {S.}~\bibnamefont {Vaidya}},\ }\bibfield  {title} {\enquote {\bibinfo {title}
  {Lectures on fuzzy and fuzzy {SUSY} physics},}\ }\href@noop {} {\  (\bibinfo
  {year} {2005})},\ \Eprint {http://arxiv.org/abs/hep-th/0511114}
  {hep-th/0511114} \BibitemShut {NoStop}%
\bibitem [{\citenamefont {Berezin}(1975)}]{Berezin:1974du}%
  \BibitemOpen
  \bibfield  {author} {\bibinfo {author} {\bibfnamefont {F.~A.}\ \bibnamefont
  {Berezin}},\ }\bibfield  {title} {\enquote {\bibinfo {title} {General concept
  of quantization},}\ }\href@noop {} {\bibfield  {journal} {\bibinfo  {journal}
  {Commun. Math. Phys.}\ }\textbf {\bibinfo {volume} {40}},\ \bibinfo {pages}
  {153--174} (\bibinfo {year} {1975})}\BibitemShut {NoStop}%
\bibitem [{\citenamefont {Hoppe}(1982)}]{Hoppe:1982}%
  \BibitemOpen
  \bibfield  {author} {\bibinfo {author} {\bibfnamefont {J.}~\bibnamefont
  {Hoppe}},\ }\emph {\bibinfo {title} {Quantum Theory of a Massless
  Relativistic Surface and a Two Dimensional Bound State Problem}},\ \href@noop
  {} {Ph.D. thesis},\ \bibinfo  {school} {MIT} (\bibinfo {year}
  {1982})\BibitemShut {NoStop}%
\bibitem [{\citenamefont {Madore}(1992)}]{Madore:1991bw}%
  \BibitemOpen
  \bibfield  {author} {\bibinfo {author} {\bibfnamefont {J.}~\bibnamefont
  {Madore}},\ }\bibfield  {title} {\enquote {\bibinfo {title} {The fuzzy
  sphere},}\ }\href@noop {} {\bibfield  {journal} {\bibinfo  {journal} {Class.
  Quant. Grav.}\ }\textbf {\bibinfo {volume} {9}},\ \bibinfo {pages} {69--88}
  (\bibinfo {year} {1992})}\BibitemShut {NoStop}%
\bibitem [{\citenamefont {Grosse}\ \emph {et~al.}(1996)\citenamefont {Grosse},
  \citenamefont {Klim\v{c}\'{\i}k},\ and\ \citenamefont
  {Pre$\check{\mathrm{s}}$najder}}]{Grosse:1995jt}%
  \BibitemOpen
  \bibfield  {author} {\bibinfo {author} {\bibfnamefont {H.}~\bibnamefont
  {Grosse}}, \bibinfo {author} {\bibfnamefont {C.}~\bibnamefont
  {Klim\v{c}\'{\i}k}}, \ and\ \bibinfo {author} {\bibfnamefont
  {P.}~\bibnamefont {Pre$\check{\mathrm{s}}$najder}},\ }\bibfield  {title}
  {\enquote {\bibinfo {title} {Topologically nontrivial field configurations in
  noncommutative geometry},}\ }\href@noop {} {\bibfield  {journal} {\bibinfo
  {journal} {Commun. Math. Phys.}\ }\textbf {\bibinfo {volume} {178}},\
  \bibinfo {pages} {507--526} (\bibinfo {year} {1996})},\ \Eprint
  {http://arxiv.org/abs/hep-th/9510083} {hep-th/9510083} \BibitemShut {NoStop}%
\bibitem [{\citenamefont {Dolan}\ \emph {et~al.}(2004)\citenamefont {Dolan},
  \citenamefont {O'Connor},\ and\ \citenamefont
  {Pre$\check{\mathrm{s}}$najder}}]{Dolan:2003th}%
  \BibitemOpen
  \bibfield  {author} {\bibinfo {author} {\bibfnamefont {Brian~P.}\
  \bibnamefont {Dolan}}, \bibinfo {author} {\bibfnamefont {Denjoe}\
  \bibnamefont {O'Connor}}, \ and\ \bibinfo {author} {\bibfnamefont {Peter}\
  \bibnamefont {Pre$\check{\mathrm{s}}$najder}},\ }\bibfield  {title} {\enquote
  {\bibinfo {title} {Fuzzy complex quadrics and spheres},}\ }\href@noop {}
  {\bibfield  {journal} {\bibinfo  {journal} {JHEP}\ }\textbf {\bibinfo
  {volume} {02}},\ \bibinfo {pages} {055} (\bibinfo {year} {2004})},\ \Eprint
  {http://arxiv.org/abs/hep-th/0312190} {hep-th/0312190} \BibitemShut {NoStop}%
\bibitem [{\citenamefont {Balachandran}\ \emph {et~al.}(2002)\citenamefont
  {Balachandran}, \citenamefont {Dolan}, \citenamefont {Lee}, \citenamefont
  {Martin},\ and\ \citenamefont {O'Connor}}]{Balachandran:2001dd}%
  \BibitemOpen
  \bibfield  {author} {\bibinfo {author} {\bibfnamefont {A.~P.}\ \bibnamefont
  {Balachandran}}, \bibinfo {author} {\bibfnamefont {Brian~P.}\ \bibnamefont
  {Dolan}}, \bibinfo {author} {\bibfnamefont {Joo-Han}\ \bibnamefont {Lee}},
  \bibinfo {author} {\bibfnamefont {X.}~\bibnamefont {Martin}}, \ and\ \bibinfo
  {author} {\bibfnamefont {Denjoe}\ \bibnamefont {O'Connor}},\ }\bibfield
  {title} {\enquote {\bibinfo {title} {Fuzzy complex projective spaces and
  their star-products},}\ }\href@noop {} {\bibfield  {journal} {\bibinfo
  {journal} {J. Geom. Phys.}\ }\textbf {\bibinfo {volume} {43}},\ \bibinfo
  {pages} {184--204} (\bibinfo {year} {2002})},\ \Eprint
  {http://arxiv.org/abs/hep-th/0107099} {hep-th/0107099} \BibitemShut {NoStop}%
\bibitem [{\citenamefont {Dolan}\ \emph {et~al.}(2007)\citenamefont {Dolan},
  \citenamefont {Huet}, \citenamefont {Murray},\ and\ \citenamefont
  {O'Connor}}]{Dolan:2006tx}%
  \BibitemOpen
  \bibfield  {author} {\bibinfo {author} {\bibfnamefont {Brian~P.}\
  \bibnamefont {Dolan}}, \bibinfo {author} {\bibfnamefont {Idrish}\
  \bibnamefont {Huet}}, \bibinfo {author} {\bibfnamefont {Se\'an}\ \bibnamefont
  {Murray}}, \ and\ \bibinfo {author} {\bibfnamefont {Denjoe}\ \bibnamefont
  {O'Connor}},\ }\bibfield  {title} {\enquote {\bibinfo {title} {Noncommutative
  vector bundles over fuzzy $\mathbb{CP}^n$ and their covariant derivatives},}\
  }\href@noop {} {\bibfield  {journal} {\bibinfo  {journal} {JHEP}\ }\textbf
  {\bibinfo {volume} {07}},\ \bibinfo {pages} {007} (\bibinfo {year} {2007})},\
  \Eprint {http://arxiv.org/abs/hep-th/0611209} {hep-th/0611209} \BibitemShut
  {NoStop}%
\bibitem [{\citenamefont {Dolan}\ \emph {et~al.}(2008)\citenamefont {Dolan},
  \citenamefont {Huet}, \citenamefont {Murray},\ and\ \citenamefont
  {O'Connor}}]{Dolan:2007}%
  \BibitemOpen
  \bibfield  {author} {\bibinfo {author} {\bibfnamefont {Brian~P.}\
  \bibnamefont {Dolan}}, \bibinfo {author} {\bibfnamefont {Idrish}\
  \bibnamefont {Huet}}, \bibinfo {author} {\bibfnamefont {Se\'an}\ \bibnamefont
  {Murray}}, \ and\ \bibinfo {author} {\bibfnamefont {Denjoe}\ \bibnamefont
  {O'Connor}},\ }\bibfield  {title} {\enquote {\bibinfo {title} {{A universal
  Dirac operator and noncommutative spin bundles over fuzzy complex projective
  spaces}},}\ }\href {\doibase 10.1088/1126-6708/2008/03/029} {\bibfield
  {journal} {\bibinfo  {journal} {JHEP}\ }\textbf {\bibinfo {volume} {03}},\
  \bibinfo {pages} {029} (\bibinfo {year} {2008})},\ \Eprint
  {http://arxiv.org/abs/0711.1347} {arXiv:0711.1347 [hep-th]} \BibitemShut
  {NoStop}%
\bibitem [{\citenamefont {Dolan}\ and\ \citenamefont
  {Jahn}(2003)}]{Dolan:2001mi}%
  \BibitemOpen
  \bibfield  {author} {\bibinfo {author} {\bibfnamefont {Brian~P.}\
  \bibnamefont {Dolan}}\ and\ \bibinfo {author} {\bibfnamefont {Oliver}\
  \bibnamefont {Jahn}},\ }\bibfield  {title} {\enquote {\bibinfo {title} {Fuzzy
  complex {G}rassmannian spaces and their star products},}\ }\href@noop {}
  {\bibfield  {journal} {\bibinfo  {journal} {Int. J. Mod. Phys.}\ }\textbf
  {\bibinfo {volume} {A18}},\ \bibinfo {pages} {1935--1958} (\bibinfo {year}
  {2003})},\ \Eprint {http://arxiv.org/abs/hep-th/0111020} {hep-th/0111020}
  \BibitemShut {NoStop}%
\bibitem [{\citenamefont {Murray}\ and\ \citenamefont
  {S\"{a}mann}(2008)}]{Murray:2006pi}%
  \BibitemOpen
  \bibfield  {author} {\bibinfo {author} {\bibfnamefont {Se\'{a}n}\
  \bibnamefont {Murray}}\ and\ \bibinfo {author} {\bibfnamefont {Christian}\
  \bibnamefont {S\"{a}mann}},\ }\bibfield  {title} {\enquote {\bibinfo {title}
  {Quantization of flag manifolds and their supersymmetric extensions},}\
  }\href@noop {} {\bibfield  {journal} {\bibinfo  {journal} {ATMP}\ }\textbf
  {\bibinfo {volume} {12}} (\bibinfo {year} {2008})},\ \Eprint
  {http://arxiv.org/abs/hep-th/0611328} {hep-th/0611328} \BibitemShut {NoStop}%
\bibitem [{\citenamefont {Arnlind}\ \emph {et~al.}(2009)\citenamefont
  {Arnlind}, \citenamefont {Bordemann}, \citenamefont {Hofer}, \citenamefont
  {Hoppe},\ and\ \citenamefont {Shimada}}]{Arnlind:2006ux}%
  \BibitemOpen
  \bibfield  {author} {\bibinfo {author} {\bibfnamefont {J.}~\bibnamefont
  {Arnlind}}, \bibinfo {author} {\bibfnamefont {M.}~\bibnamefont {Bordemann}},
  \bibinfo {author} {\bibfnamefont {L.}~\bibnamefont {Hofer}}, \bibinfo
  {author} {\bibfnamefont {J.}~\bibnamefont {Hoppe}}, \ and\ \bibinfo {author}
  {\bibfnamefont {Hidehiko}\ \bibnamefont {Shimada}},\ }\bibfield  {title}
  {\enquote {\bibinfo {title} {{Fuzzy Riemann surfaces}},}\ }\href {\doibase
  10.1088/1126-6708/2009/06/047} {\bibfield  {journal} {\bibinfo  {journal}
  {JHEP}\ }\textbf {\bibinfo {volume} {06}},\ \bibinfo {pages} {047} (\bibinfo
  {year} {2009})},\ \Eprint {http://arxiv.org/abs/hep-th/0602290}
  {arXiv:hep-th/0602290} \BibitemShut {NoStop}%
\bibitem [{\citenamefont {Lizzi}\ \emph {et~al.}(2003)\citenamefont {Lizzi},
  \citenamefont {Vitale},\ and\ \citenamefont {Zampini}}]{Lizzi:2003ru}%
  \BibitemOpen
  \bibfield  {author} {\bibinfo {author} {\bibfnamefont {F.}~\bibnamefont
  {Lizzi}}, \bibinfo {author} {\bibfnamefont {P.}~\bibnamefont {Vitale}}, \
  and\ \bibinfo {author} {\bibfnamefont {A.}~\bibnamefont {Zampini}},\
  }\bibfield  {title} {\enquote {\bibinfo {title} {{The fuzzy disc}},}\
  }\href@noop {} {\bibfield  {journal} {\bibinfo  {journal} {JHEP}\ }\textbf
  {\bibinfo {volume} {08}},\ \bibinfo {pages} {057} (\bibinfo {year} {2003})},\
  \Eprint {http://arxiv.org/abs/hep-th/0306247} {arXiv:hep-th/0306247}
  \BibitemShut {NoStop}%
\bibitem [{\citenamefont {Lizzi}\ \emph {et~al.}(2005)\citenamefont {Lizzi},
  \citenamefont {Vitale},\ and\ \citenamefont {Zampini}}]{Lizzi:2005zx}%
  \BibitemOpen
  \bibfield  {author} {\bibinfo {author} {\bibfnamefont {Fedele}\ \bibnamefont
  {Lizzi}}, \bibinfo {author} {\bibfnamefont {Patrizia}\ \bibnamefont
  {Vitale}}, \ and\ \bibinfo {author} {\bibfnamefont {Alessandro}\ \bibnamefont
  {Zampini}},\ }\bibfield  {title} {\enquote {\bibinfo {title} {{The beat of a
  fuzzy drum: fuzzy Bessel functions for the disc}},}\ }\href@noop {}
  {\bibfield  {journal} {\bibinfo  {journal} {JHEP}\ }\textbf {\bibinfo
  {volume} {09}},\ \bibinfo {pages} {080} (\bibinfo {year} {2005})},\ \Eprint
  {http://arxiv.org/abs/hep-th/0506008} {arXiv:hep-th/0506008} \BibitemShut
  {NoStop}%
\bibitem [{\citenamefont {Scholtz}\ \emph {et~al.}(2007)\citenamefont
  {Scholtz}, \citenamefont {Chakraborty}, \citenamefont {Govaerts},\ and\
  \citenamefont {Vaidya}}]{Scholtz:2007ig}%
  \BibitemOpen
  \bibfield  {author} {\bibinfo {author} {\bibfnamefont {F.~G.}\ \bibnamefont
  {Scholtz}}, \bibinfo {author} {\bibfnamefont {B.}~\bibnamefont
  {Chakraborty}}, \bibinfo {author} {\bibfnamefont {J.}~\bibnamefont
  {Govaerts}}, \ and\ \bibinfo {author} {\bibfnamefont {S.}~\bibnamefont
  {Vaidya}},\ }\bibfield  {title} {\enquote {\bibinfo {title} {{Spectrum of the
  non-commutative spherical well}},}\ }\href {\doibase
  10.1088/1751-8113/40/48/019} {\bibfield  {journal} {\bibinfo  {journal} {J.
  Phys.}\ }\textbf {\bibinfo {volume} {A40}},\ \bibinfo {pages} {14581--14592}
  (\bibinfo {year} {2007})},\ \Eprint {http://arxiv.org/abs/0709.3357}
  {arXiv:0709.3357 [hep-th]} \BibitemShut {NoStop}%
\bibitem [{\citenamefont {Hammou}\ \emph {et~al.}(2002)\citenamefont {Hammou},
  \citenamefont {Lagraa},\ and\ \citenamefont
  {Sheikh-Jabbari}}]{Hammou:2001cc}%
  \BibitemOpen
  \bibfield  {author} {\bibinfo {author} {\bibfnamefont {A.~B.}\ \bibnamefont
  {Hammou}}, \bibinfo {author} {\bibfnamefont {M.}~\bibnamefont {Lagraa}}, \
  and\ \bibinfo {author} {\bibfnamefont {M.~M.}\ \bibnamefont
  {Sheikh-Jabbari}},\ }\bibfield  {title} {\enquote {\bibinfo {title}
  {{Coherent state induced star-product on R(lambda)**3 and the fuzzy
  sphere}},}\ }\href {\doibase 10.1103/PhysRevD.66.025025} {\bibfield
  {journal} {\bibinfo  {journal} {Phys. Rev.}\ }\textbf {\bibinfo {volume}
  {D66}},\ \bibinfo {pages} {025025} (\bibinfo {year} {2002})},\ \Eprint
  {http://arxiv.org/abs/hep-th/0110291} {arXiv:hep-th/0110291} \BibitemShut
  {NoStop}%
\bibitem [{\citenamefont {Moreno}(2005)}]{Moreno:2005cn}%
  \BibitemOpen
  \bibfield  {author} {\bibinfo {author} {\bibfnamefont {E.~F.}\ \bibnamefont
  {Moreno}},\ }\bibfield  {title} {\enquote {\bibinfo {title} {{Spherically
  symmetric monopoles in noncommutative space}},}\ }\href {\doibase
  10.1103/PhysRevD.72.045001} {\bibfield  {journal} {\bibinfo  {journal} {Phys.
  Rev.}\ }\textbf {\bibinfo {volume} {D72}},\ \bibinfo {pages} {045001}
  (\bibinfo {year} {2005})},\ \Eprint {http://arxiv.org/abs/hep-th/0506134}
  {arXiv:hep-th/0506134} \BibitemShut {NoStop}%
\bibitem [{\citenamefont {Buric}\ and\ \citenamefont
  {Madore}(2008)}]{Buric:2008th}%
  \BibitemOpen
  \bibfield  {author} {\bibinfo {author} {\bibfnamefont {Maja}\ \bibnamefont
  {Buric}}\ and\ \bibinfo {author} {\bibfnamefont {John}\ \bibnamefont
  {Madore}},\ }\bibfield  {title} {\enquote {\bibinfo {title} {{Spherically
  Symmetric Noncommutative Space: d = 4}},}\ }\href {\doibase
  10.1140/epjc/s10052-008-0748-6} {\bibfield  {journal} {\bibinfo  {journal}
  {Eur. Phys. J.}\ }\textbf {\bibinfo {volume} {C58}},\ \bibinfo {pages}
  {347--353} (\bibinfo {year} {2008})},\ \Eprint
  {http://arxiv.org/abs/0807.0960} {arXiv:0807.0960 [hep-th]} \BibitemShut
  {NoStop}%
\bibitem [{\citenamefont {Snyder}(1947)}]{Snyder:1946qz}%
  \BibitemOpen
  \bibfield  {author} {\bibinfo {author} {\bibfnamefont {Hartland~S.}\
  \bibnamefont {Snyder}},\ }\bibfield  {title} {\enquote {\bibinfo {title}
  {{Quantized space-time}},}\ }\href {\doibase 10.1103/PhysRev.71.38}
  {\bibfield  {journal} {\bibinfo  {journal} {Phys. Rev.}\ }\textbf {\bibinfo
  {volume} {71}},\ \bibinfo {pages} {38--41} (\bibinfo {year}
  {1947})}\BibitemShut {NoStop}%
\bibitem [{Note1()}]{Note1}%
  \BibitemOpen
  \bibinfo {note} {Repeated indices are summed over.}\BibitemShut {Stop}%
\bibitem [{\citenamefont {Wybourne}(1974)}]{Wybourne}%
  \BibitemOpen
  \bibfield  {author} {\bibinfo {author} {\bibfnamefont {Brian}\ \bibnamefont
  {Wybourne}},\ }\href@noop {} {\emph {\bibinfo {title} {Classical Groups for
  Physicists}}}\ (\bibinfo  {publisher} {John Wiley \& Sons},\ \bibinfo {year}
  {1974})\BibitemShut {NoStop}%
\bibitem [{Note2()}]{Note2}%
  \BibitemOpen
  \bibinfo {note} {In fact, the algebra relations (\ref {JJcommutator}) and
  (\ref {JXcommutator}) alone imply that the operators $\protect \mathaccentV
  {hat}05E{J}^2$, $\protect \mathaccentV {hat}05E{J}_3$ and $\protect
  \mathaccentV {hat}05E{J}\cdot \protect \mathaccentV {hat}05E{X}=\protect
  \mathaccentV {hat}05E{X}\cdot \protect \mathaccentV {hat}05E{J}$ mutually
  commute and so their eigenvalues can be used to label the Hilbert space.
  Using the additional commutator (\ref {genXXcommutator}), we find that
  $\protect \mathaccentV {hat}05E{X}^2$ also mutually commutes and its
  eigenvalue is also a label. The two Casimirs are formed from $\protect
  \mathaccentV {hat}05E{J}^2$, $\protect \mathaccentV {hat}05E{J}\cdot \protect
  \mathaccentV {hat}05E{X}$ and $\protect \mathaccentV
  {hat}05E{X}^2$.}\BibitemShut {Stop}%
\bibitem [{Note3()}]{Note3}%
  \BibitemOpen
  \bibinfo {note} {For generality, we allow spinor
  representations.}\BibitemShut {Stop}%
\bibitem [{Note4()}]{Note4}%
  \BibitemOpen
  \bibinfo {note} {In Snyder's original construction, time is included and the
  noncommuting coordinates are covariant under Lorentz transformation. He also
  extends this algebra by adding commuting momentum operators.}\BibitemShut
  {Stop}%
\bibitem [{Note5()}]{Note5}%
  \BibitemOpen
  \bibinfo {note} {For a discusion of the Euclidean group in the context of
  magnetic charge quantization, see \cite
  {Lipkin:1969ck,Peshkin:1971bg}.}\BibitemShut {Stop}%
\bibitem [{\citenamefont {Adams}\ \emph
  {et~al.}(1988{\natexlab{a}})\citenamefont {Adams}, \citenamefont
  {\v{C}\'{\i}\v{z}ek},\ and\ \citenamefont {Paldus}}]{Adams19881}%
  \BibitemOpen
  \bibfield  {author} {\bibinfo {author} {\bibfnamefont {B.G.}\ \bibnamefont
  {Adams}}, \bibinfo {author} {\bibfnamefont {J.}~\bibnamefont
  {\v{C}\'{\i}\v{z}ek}}, \ and\ \bibinfo {author} {\bibfnamefont
  {J.}~\bibnamefont {Paldus}},\ }\bibfield  {title} {\enquote {\bibinfo {title}
  {Lie algebraic methods and their applications to simple quantum systems},}\
  }in\ \href@noop {} {\emph {\bibinfo {booktitle} {Advances in Quantum
  Chemistry}}},\ Vol.~\bibinfo {volume} {19},\ \bibinfo {editor} {edited by\
  \bibinfo {editor} {\bibnamefont {{Per-Olov L\"owdin}}}}\ (\bibinfo
  {publisher} {Academic Press},\ \bibinfo {year} {1988})\ pp.\ \bibinfo {pages}
  {1 -- 85},\ \bibinfo {note} {reprinted \cite{Bohm1988}}\BibitemShut {NoStop}%
\bibitem [{\citenamefont {Adams}\ \emph
  {et~al.}(1988{\natexlab{b}})\citenamefont {Adams}, \citenamefont
  {\v{C}\'{\i}\v{z}ek},\ and\ \citenamefont {Paldus}}]{Bohm1988}%
  \BibitemOpen
  \bibfield  {author} {\bibinfo {author} {\bibfnamefont {B.G.}\ \bibnamefont
  {Adams}}, \bibinfo {author} {\bibfnamefont {J.}~\bibnamefont
  {\v{C}\'{\i}\v{z}ek}}, \ and\ \bibinfo {author} {\bibfnamefont
  {J.}~\bibnamefont {Paldus}},\ }\bibfield  {title} {\enquote {\bibinfo {title}
  {Lie algebraic methods and their applications to simple quantum systems},}\
  }in\ \href@noop {} {\emph {\bibinfo {booktitle} {Dynamical groups and
  spectrum generating algebras}}},\ Vol.~\bibinfo {volume} {1},\ \bibinfo
  {editor} {edited by\ \bibinfo {editor} {\bibnamefont {{A. B\"ohm and Y.
  Ne'eman and A. Barut}}}}\ (\bibinfo  {publisher} {World Scientific
  Publishing},\ \bibinfo {year} {1988})\ pp.\ \bibinfo {pages}
  {103--208}\BibitemShut {NoStop}%
\bibitem [{Note6()}]{Note6}%
  \BibitemOpen
  \bibinfo {note} {Since the algebra is isomorphic to the standard similarity
  algebra, we can introduce momentum operators $\protect \mathaccentV
  {hat}05E{p}_i$ and write the coordinate operators in term of oscillators
  $a^i=\protect \frac {1}{\protect \sqrt {2}}(\protect \mathaccentV
  {hat}05E{x}_i+i\protect \mathaccentV {hat}05E{p}_i)$, $a^\dagger _i=\protect
  \frac {1}{\protect \sqrt {2}}(\protect \mathaccentV {hat}05E{x}_i-i\protect
  \mathaccentV {hat}05E{p}_i)$, $i=1,2,3$. The angular momentum operators are
  associated to the $su(2)$ subalgebra, $\protect \mathaccentV
  {hat}05E{J}_i=-i\epsilon _{ijk}a^\dagger _ja^k$, and the Laplacian $\protect
  \mathaccentV {hat}05E{p}^2$ is invariant under the corresponding subgroup.
  The realisation we have presented above however, only describes that part of
  the Hilbert space generated by the (noncommuting) coordinates.}\BibitemShut
  {Stop}%
\bibitem [{Note7()}]{Note7}%
  \BibitemOpen
  \bibinfo {note} {If one attempts to add commuting momentum operators
  $\protect \mathaccentV {hat}05E{P}_i$ to the Snyder algebra, then there are
  infinitely many ways to choose the $[\protect \mathaccentV
  {hat}05E{X}_i,\protect \tmspace +\thinmuskip {.1667em}\protect \mathaccentV
  {hat}05E{P}_j]$ and still have a consistent algebra \cite {Battisti:2008xy}.
  Here, we have modified the Snyder algebra such that it is isomorphic to
  $e(3)$. We then have a natural choice for this commutator: $[\protect
  \mathaccentV {hat}05E{X}_i,\protect \tmspace +\thinmuskip {.1667em}\protect
  \mathaccentV {hat}05E{P}_j]=i\delta _{ij}+i\protect \frac {\beta }{2}\epsilon
  _{ijk}\protect \mathaccentV {hat}05E{P}_k$.}\BibitemShut {Stop}%
\bibitem [{Note8()}]{Note8}%
  \BibitemOpen
  \bibinfo {note} {Note that $\protect \frac {j_l(kr)}{r^l}$ is expressable as
  a power series that contains only even (non-negative) powers of
  $r$.}\BibitemShut {Stop}%
\bibitem [{\citenamefont {Bi\`evre}(1989)}]{bievre:1401}%
  \BibitemOpen
  \bibfield  {author} {\bibinfo {author} {\bibfnamefont {S.~De}\ \bibnamefont
  {Bi\`evre}},\ }\bibfield  {title} {\enquote {\bibinfo {title} {Coherent
  states over symplectic homogeneous spaces},}\ }\href {\doibase
  10.1063/1.528321} {\bibfield  {journal} {\bibinfo  {journal} {J. Math. Phys}\
  }\textbf {\bibinfo {volume} {30}},\ \bibinfo {pages} {1401--1407} (\bibinfo
  {year} {1989})}\BibitemShut {NoStop}%
\bibitem [{\citenamefont {Isham}\ and\ \citenamefont
  {Klauder}(1991)}]{isham:607}%
  \BibitemOpen
  \bibfield  {author} {\bibinfo {author} {\bibfnamefont {C.~J.}\ \bibnamefont
  {Isham}}\ and\ \bibinfo {author} {\bibfnamefont {J.~R.}\ \bibnamefont
  {Klauder}},\ }\bibfield  {title} {\enquote {\bibinfo {title} {Coherent states
  for n-dimensional {Euclidean groups E(n)} and their application},}\ }\href
  {\doibase 10.1063/1.529402} {\bibfield  {journal} {\bibinfo  {journal} {J.
  Math. Phys}\ }\textbf {\bibinfo {volume} {32}},\ \bibinfo {pages} {607--620}
  (\bibinfo {year} {1991})}\BibitemShut {NoStop}%
\bibitem [{\citenamefont {Lipkin}\ \emph {et~al.}(1969)\citenamefont {Lipkin},
  \citenamefont {Weisberger},\ and\ \citenamefont {Peshkin}}]{Lipkin:1969ck}%
  \BibitemOpen
  \bibfield  {author} {\bibinfo {author} {\bibfnamefont {H.~J.}\ \bibnamefont
  {Lipkin}}, \bibinfo {author} {\bibfnamefont {W.~I.}\ \bibnamefont
  {Weisberger}}, \ and\ \bibinfo {author} {\bibfnamefont {M.}~\bibnamefont
  {Peshkin}},\ }\bibfield  {title} {\enquote {\bibinfo {title} {{Magnetic
  charge quantization and angular momentum}},}\ }\href {\doibase
  10.1016/0003-4916(69)90279-6} {\bibfield  {journal} {\bibinfo  {journal}
  {Annals Phys.}\ }\textbf {\bibinfo {volume} {53}},\ \bibinfo {pages}
  {203--214} (\bibinfo {year} {1969})}\BibitemShut {NoStop}%
\bibitem [{\citenamefont {Peshkin}(1971)}]{Peshkin:1971bg}%
  \BibitemOpen
  \bibfield  {author} {\bibinfo {author} {\bibfnamefont {M.}~\bibnamefont
  {Peshkin}},\ }\bibfield  {title} {\enquote {\bibinfo {title} {{Elementary
  algebra of the Euclidean group, with application to magnetic charge
  quantization}},}\ }\href {\doibase 10.1016/0003-4916(71)90069-8} {\bibfield
  {journal} {\bibinfo  {journal} {Annals Phys.}\ }\textbf {\bibinfo {volume}
  {66}},\ \bibinfo {pages} {542--547} (\bibinfo {year} {1971})}\BibitemShut
  {NoStop}%
\bibitem [{\citenamefont {Battisti}\ and\ \citenamefont
  {Meljanac}(2009)}]{Battisti:2008xy}%
  \BibitemOpen
  \bibfield  {author} {\bibinfo {author} {\bibfnamefont {Marco~Valerio}\
  \bibnamefont {Battisti}}\ and\ \bibinfo {author} {\bibfnamefont {Stjepan}\
  \bibnamefont {Meljanac}},\ }\bibfield  {title} {\enquote {\bibinfo {title}
  {{Modification of Heisenberg uncertainty relations in non- commutative Synder
  space-time geometry}},}\ }\href {\doibase 10.1103/PhysRevD.79.067505}
  {\bibfield  {journal} {\bibinfo  {journal} {Phys. Rev.}\ }\textbf {\bibinfo
  {volume} {D79}},\ \bibinfo {pages} {067505} (\bibinfo {year} {2009})},\
  \Eprint {http://arxiv.org/abs/0812.3755} {arXiv:0812.3755 [hep-th]}
  \BibitemShut {NoStop}%
\end{thebibliography}%

\end{document}